\documentclass[aps,prx,twocolumn,superscriptaddress,showpacs,floatfix]{revtex4-2}
\usepackage{comment}
\usepackage{physics}
\usepackage{amsmath,amsthm,amssymb,amsfonts, color, comment, graphicx, slashed}%, %fancyhdr, environ}
\usepackage{xcolor}
\usepackage{float}
\usepackage{graphicx}
\usepackage{dcolumn}% Align table columns on decimal point
\usepackage{bm}% bold math
\UseRawInputEncoding 
\usepackage{dsfont}
\usepackage{hyperref}% add hypertext capabilities

\newcommand{\bea}{\begin{eqnarray}}
\newcommand{\eea}{\end{eqnarray}}

\newcommand{\bsigma}{\boldsymbol{\sigma}}
\newcommand{\bS}{\mathbf{S}}

\newcommand{\br}{\mathbf{r}}

\newcommand{\bGa}{\boldsymbol{\Gamma}}

\def\kcso{K$_2$Co(SeO$_3$)$_2$}
\def\ncpo{Na$_2$BaCo(PO$_4$)$_2$}
\def\rcso{Rb$_2$Co(SeO$_3$)$_2$}
\def\ybzn{YbZn$_2$GaO$_5$}
\def\ybmg{YbMgGaO$_4$}
\def\cmao{CeMgAl$_{11}$O$_{19}$}
\def\bcso{Ba$_3$CoSb$_2$O$_9$}

\newcommand{\be}{\begin{equation}}
\newcommand{\ee}{\end{equation}}
\newcommand{\bk}{{{\bf{k}}}}

\newcommand{\bK}{{{\bf{K}}}}

\newcommand{\bQ}{{{\bf{Q}}}}
\newcommand{\bM}{{{\bf{M}}}}

\newcommand{\beal}{\begin{align}}
\newcommand{\eeal}{\end{align}}
\newcommand{\ra}{\rangle}
\newcommand{\la}{\langle}

\newcommand{\dg}{{\dagger}}
\newcommand{\pdg}{{\phantom\dagger}}

\renewcommand{\vec}[1]{\mathbf{#1}}

\newcommand{\btjstrw}{\mathrel{{\rotatebox[origin=c]{90}
{$\bowtie$}}\kern-0.18em\raisebox{-.95ex}{$\bullet$}
\kern-0.5em\raisebox{.97ex}{$\bullet$}
\kern-1.12em\raisebox{.97ex}{$\bullet$}
\kern-0.52em\raisebox{-.95ex}{$\bullet$}}}

\newcommand{\btjnbrR}{{\mathrel{\rotatebox[origin=c]{90}
{$\bowtie$}}\kern-0.22em\raisebox{.9ex}{$\bullet$}
\kern-1.em\raisebox{-.8ex}{$\bullet$}}}
\newcommand{\btjnbrL}{{\mathrel{\rotatebox[origin=c]{90}
{$\bowtie$}}\kern-0.22em\raisebox{-.8ex}{$\bullet$}
\kern-1.em\raisebox{+.9ex}{$\bullet$}}}

\def\a{\alpha}
\def\b{\beta}
\def\c{\chi}
\def\d{\delta}

\def\m{\mu}
\def\n{\nu}

\def\r{\rho}
\def\s{\sigma}
\def\t{\tau}

\def\w{\omega}

\def\D{\Delta}

\def\W{\Omega}

\def\mg{{\mathcal{G}}}
\def\mh{{\mathcal{H}}}

\def\mj{{\mathcal{J}}}

\def\mo{{\mathcal{O}}}

\def\ms{{\mathcal{S}}}

\def\ua{\uparrow}
\def\da{\downarrow}
\usepackage{graphicx}                                  
\usepackage[caption=false]{subfig}

\newcolumntype{P}[1]{>{\centering\arraybackslash}p{#1}}

\begin{document}
\title{Modified large-$N$ approach to gapless spin liquids, magnetic orders, and dynamics: Application to
triangular lattice antiferromagnets}
\author{Anjishnu Bose}
\email{anjishnu.bose@mail.utoronto.ca}
\affiliation{Department of Physics, University of Toronto, 60 St. George Street, Toronto, ON, M5S 1A7 Canada}
\author{Kathleen Hart}
\affiliation{Department of Physics, University of Toronto, 60 St. George Street, Toronto, ON, M5S 1A7 Canada}
\author{Ruairidh Sutcliffe}
\affiliation{Department of Physics, University of Toronto, 60 St. George Street, Toronto, ON, M5S 1A7 Canada}
\author{Arun Paramekanti}
\email{arun.paramekanti@utoronto.ca}
\affiliation{Department of Physics, University of Toronto, 60 St. George Street, Toronto, ON, M5S 1A7 Canada}
\date{\today}
\begin{abstract}
Recent work has shown that the triangular lattice spin-$1/2$ $J_1$-$J_2$ Heisenberg and XXZ antiferromagnets may exhibit coplanar or supersolid orders proximate to a gapless Dirac spin liquid phase. We explore a distinct $SU(2N)\!\!\times\!\!SU(M)$
fermionic parton approach, complemented by variational Monte Carlo calculations for the spin-$1/2$ model, 
to study the phase diagram of these models. We also calculate their dynamical spin response including parton interactions within a random phase approximation, and discuss implications for neutron scattering on 
triangular lattice cobaltates Ba$_3$CoSb$_2$O$_9$, Na$_2$BaCo(PO$_4$)$_2$, K$_2$Co(SeO$_3$)$_2$, Rb$_2$Co(SeO$_3$)$_2$, and 
Yb-based magnet {KYbSe$_2$}.
\end{abstract}
\maketitle

\section{Introduction}

Understanding the quantum spin dynamics of strongly frustrated spin-$1/2$ systems remains an important problem in magnetism. 
While sharp magnon modes in an ordered magnet are correctly described by linear spin-wave theory (LSWT) \cite{AuerbachBook,YosidaBook}, 
incorporating 
magnon interactions to go beyond LSWT \cite{Oguchi1960_magnoninteraction,Harris1971_magnoninteraction,magnon1992}
is necessary to capture quantitative renormalizations of the spin wave dispersion as well as the possible breakdown of magnons
\cite{Chubukov_largeS1994,Chernyshev2006_NonCollinearDecay,Chernyshev2009_TriangularAFM,Mourigal2013_Decay,Maksimov2016_Decay,spinwaveXXZ_chernyshev_PRB2022,Wang2024}.
LSWT and the leading corrections to LWST may be formally justified within a $1/S$ expansion for large spin-$S$. However, for spin-$1/2$ systems, series expansion
techniques have shown that there are interesting features in the spin dynamics which seem beyond the purview of an 
interacting magnon approach, including unusual roton-like features in the excitation spectrum \cite{Zheng2006_PRLXXZTriangular_SeriesExpansion,Zheng2006_PRBXXZTriangular_SeriesExpansion}.
The discoveries of several highly frustrated
spin-$1/2$ magnetic materials which defy ordering down to temperatures well below the Curie-Weiss scale, i.e. with $T_N \ll \theta_{\rm CW}$,
also suggests that an alternative framework to describe their quantum spin dynamics, perhaps in terms of fractionalized parton degrees of freedom,
might provide a useful perspective. 

A well-studied route to fractionalization is to express spin-$1/2$ degrees of freedom in terms of 
constrained Schwinger bosons via 
\begin{equation}
    \vec S = \frac{1}{2} b^\dg_{\alpha} \bsigma_{\alpha\beta} b^\pdg_\beta;~~b^\dg_{\alpha} b^\pdg_\alpha=1\,
\end{equation}
with implicit sum on spinor indices.
Generalizing this formulation to $\alpha=1 \ldots N$ and replacing $\bsigma_{\alpha\beta}$ by $SU(N)$ generators leads to the $SU(N)$ Schwinger boson formulation which allows one to describe both the ordered collinear N\'eel antiferromagnet on the square lattice as well as gapped dimerized magnets in Heisenberg models within a large-$N$ framework \cite{Auerbach1988,Arovas1988_SUN,Zhang2022_SchwingerBoson,Read1989_SUN}, separated by an
unconventional quantum critical point involving partons coupled to fluctuating $U(1)$ gauge fields \cite{Senthil2004_Deconfined}.
A distinct $Sp(2N)$ generalization involves generalizing two-component bosons to $N$ flavors of two-component bosons (with an additional flavor index $\ell=1\ldots N$) \cite{LargeN_Sachdev1991, SutherlandSP2N, SuperconductivitySp2N, Martins_2002, Ran2006ContinuousQP,Lawler2008,Bernier2008}. 
The spin rotation invariant Hamiltonian is then generalized as (with implicit sums on flavor and spinor indices)
\begin{eqnarray}
   H&=& J \sum_{\la i,j\ra}  \epsilon_{\alpha\beta} b^\dg_{i\ell\alpha} b^\dg_{j\ell\beta}
\epsilon_{\mu\nu} b^\pdg_{i\ell'\mu} b^\pdg_{j\ell'\nu}
\end{eqnarray}
with the constraint $b^\dg_{i\ell\alpha} b^\pdg_{i\ell\alpha}=2 N \kappa$ at each site $i$.
The physical limit of the spin model corresponds to $N=1$ and $\kappa=1/2$, but more generally
the filling factor $\kappa$ plays the role of a `spin'. This formulation is useful to describe geometrically
frustrated magnets at large $N$, with gapped singlet ${\mathbb Z}_2$ spin liquids appearing at small spin
$\kappa < \kappa_c$ and non-collinear magnetically ordered phases appearing
beyond a critical spin value $\kappa > \kappa_c$.

In order to describe {\it gapless} spin liquid phases on the other hand, it is
expedient to use a Schwinger fermion formulation \cite{QSLgauge_Lee2014, QSLgauge_XGWenBook, QSLreview_Broholm2020, QSLreview_McQueen2021, QSLreview_Savary_2017}
\begin{equation}
    \vec S = \frac{1}{2} f^\dg_{\alpha} \bsigma_{\alpha\beta} f^\pdg_\beta;~~f^\dg_{\alpha} f^\pdg_\alpha=1
\end{equation}
The $SU(N)$ Schwinger boson approach naturally generalizes to Schwinger fermions, allowing one to describe gapless spin liquids at 
large $N$ \cite{Affleck1988_SUN,Hermele_2004_Stability,Hermele2005_ASL,Lawler2008_Hyperkagome}. However, this does not allow for magnetic order in the leading large $N$ mean field limit. With decreasing $N$, magnetic order can instead emerge from strong gauge field fluctuations necessitated by the local constraint, either via proliferation of monopoles which carry nontrivial spin quantum numbers \cite{DiracMonopoles_PRX2020, Song2019}, or through four-fermion interaction terms which become relevant as \(N\) 
is decreased, which fosters chiral symmetry breaking \cite{Hermele_2004_Stability}.
It is a challenge to study these competing effects in a controlled manner.

Motivated by this, we explore here a fermionic $SU(2N)\times SU(M)$ approach which allows one to study magnetically ordered states and gapless spin liquids on equal footing. For fixed $\kappa=M/2N$, and for $N\to\infty$ this problem reduces to a magnetic order parameter 
Yukawa-coupled to fermions, reminiscent of other large-$N$ approaches to fermionic quantum critical points \cite{Damia2019_yukawa1,Wang2020_yukawa2,Wang2020_yukawa3, PhysRevLett.134.036502}, while decreasing $N$ leads to important gauge fluctuation corrections. We thus believe this approach might allow one to study fermionic quantum critical points and to systematically incorporate the impact of gauge fluctuations to study the spin liquid to
magnetic ordering transition. We use this parton approach to obtain the mean field theory of the triangular antiferromagnet, which shows 120-degree coplanar order at large `spin' and a Dirac spin liquid at small `spin'. Appendix A and B present further details of the parton theory, 
and its path integral formulation.

We then discuss a phenomenological extension of this approach to XXZ models which break the Heisenberg symmetry, and show that it can reasonably capture the ground state phase diagram of models which have been explored using density-matrix renormalization group (DMRG) computations \cite{DMRGWhite, DMRG_Sheng, DMRG_DSL, Chernyshev2024}. We complement this with variational Monte Carlo (VMC) simulations of Gutzwiller projected Dirac wavefunctions with long range Jastrow factors which lead to results in good quantitative agreement with the DMRG results, and extending previous VMC results on the Heisenberg model \cite{J1J2_triangular, J1J2_dynamics}. Details of the
VMC are in Appendix C. 

Finally, we turn to the study of the dynamical spin response of these models within the parton approach going beyond mean field theory to include parton interactions within a random phase approximation which captures recent neutron scattering results showing magnons coexisting with continuum scattering in pseudospin-$1/2$ triangular cobaltate magnets \bcso\, \cite{bcso_dynamics, bcso_Huang_2022, bcso_nature, bcso_prl},  Na$_2$BaCo(PO$_4$)$_2$ \cite{NBCOExperiment, sheng2024continuumspinexcitationsordered, SuperSolidGang}, K$_2$Co(SeO$_3$)$_2$ \cite{Zhu_2024, chen2024phasediagramspectroscopicsignatures} and Rb$_2$Co(SeO$_3$)$_2$ \cite{PhysRevMaterials.4.084406} and Yb-based magnets such as KYbSe$_2$ \cite{Scheie2024}.

Our semi-analytical approach is argued to provide insights into results obtained by more numerically intensive many-body methods such as density-matrix renormalization group (DMRG) \cite{dmrg_schwol, dmrg_Catarina_2023}, variational Monte Carlo (VMC) \cite{becca_sorella_2017, song2024neuralquantumstatesvariational}, or pseudo-fermion functional renormalization group (pfFRG) \cite{pffrg_Bfield, pffrg_rev, pffrg_pyrocholore} techniques which are used to study the ground states and spin excitations of diverse frustrated magnets. 

\section{Parton theory} \label{methodology}

\subsection{Heisenberg models}

\begin{figure}[t]
 \centering
 \includegraphics[width=0.45\textwidth]{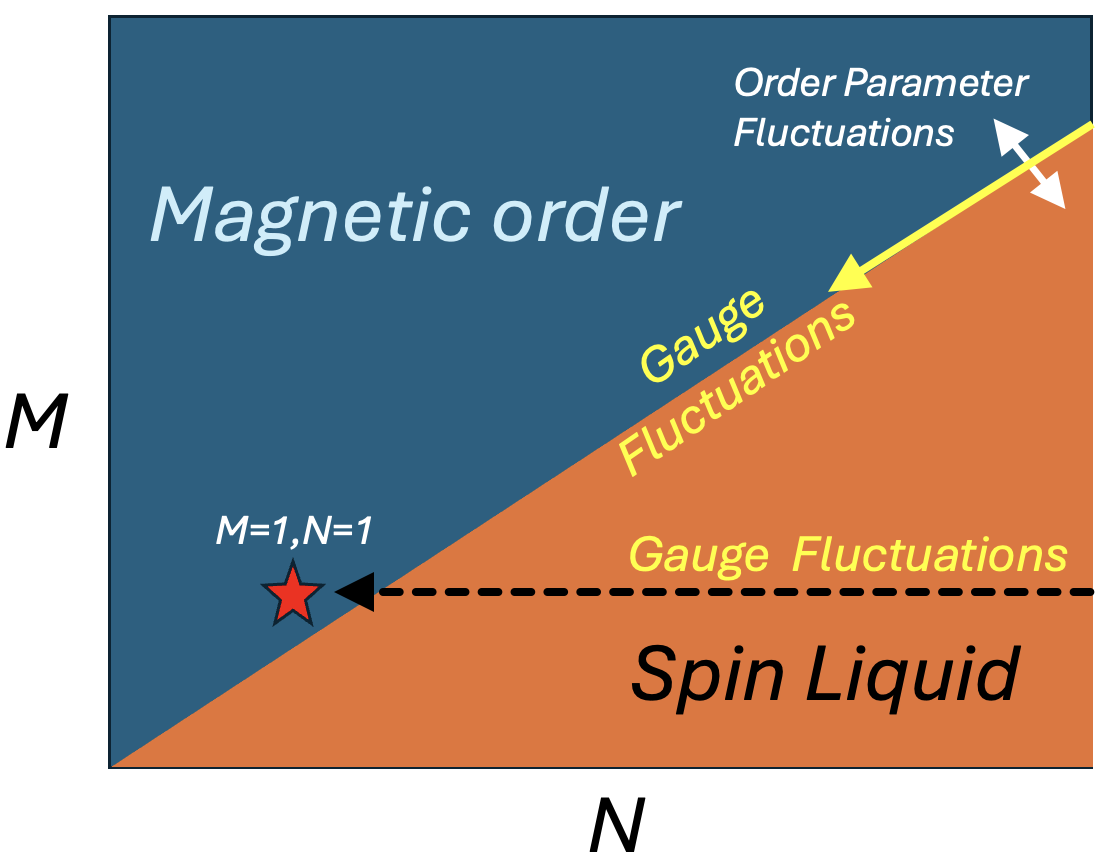}
 \caption{Scenario for the generalized $SU(2N) \times SU(M)$ parton model where large-$N$ and large-$M$ respectively favor quantum spin liquid
 (QSL) and magnetically ordered ground states. Star indicates the physical $SU(2)$ spin-$1/2$ point with $M\!=\!1,N\!=\!1$.
Trajectories of fixed $\kappa=M/2N$ correspond to lines of fixed slope, with the magnetic ordering out of the QSL occurring
at a certain $\kappa_c$. For 
 $N \to \infty$, we expect gauge fluctuations to be suppressed, so the QSL to magnetic ordering transition at $\kappa_c$
 would be driven by order parameter fluctuations. For fixed $\kappa$, decreasing $N$ leads to enhanced gauge fluctuations, so
 partons are coupled to both order parameter fluctuations and dynamical gauge fields. For tuning $N$ at
 fixed $M=1$ (dashed black line), magnetic ordering can occur due to chiral symmetry breaking or monopole
 proliferation.}
\label{fig:schematic}
\end{figure}

We begin by exploring a formal limit for Heisenberg models which describes the competition between disordered gapless spin-liquids and 
ordered symmetry-broken states for which mean-field theory is exact, a scenario which is sketched in Fig.~\ref{fig:schematic}. 
In such a limit, we expect that both spin liquid states and ordered states are energetically competitive, but all gauge fluctuations have been suppressed. Our work is inspired by the $Sp(2N)$ and large-S approaches \cite{LargeN_Sachdev1991,Chubokov_largeSHubbard} to antiferromagnets. Using fermionic partons, we introduce a flavor index \(\ell\) in the fundamental representation of \(SU(M)\), along with enlarging the \emph{physical} internal symmetry group to \(SU(2N)\). For partons at half-filling, assuming implicit summation on spinor and flavor indices, $f^{\dagger}_{i, \a, \ell}f^\pdg_{i, \a, \ell} \!=\! N M$ at each site $i$. The case of spin-\(1/2\) in \(SU(2)\) corresponds to \(N\!=\! 1\) and \(M\!=\! 1\). Now consider a Hamiltonian on bond $(i,j)$
\begin{eqnarray}
    \label{large-N,M Hamiltonian}
      \mh_{ij} &=& {\mathbb J}_{ij} f^{\dagger}_{i, \a, \ell}T^a_{\a\b}f^\pdg_{i, \b, \ell}f^{\dagger}_{j, \m, \ell'}T^a_{\m\n}f^\pdg_{j, \n, \ell'}
\end{eqnarray}
where \(T^a\) are the \(4N^2-1\) generators of \(SU(2N)\) in the fundamental representation. Using the completeness relation of the
generators, we can rewrite this in the form
\begin{eqnarray}
      \label{large-N,M Hamiltonian2}
      \mh_{ij} &=& \frac{1}{2} {\mathbb J}_{ij} f^{\dagger}_{i, \a, \ell}f^\pdg_{i, \b, \ell}f^{\dagger}_{j, \b, \ell'}f^\pdg_{j, \a, \ell'} + \mathrm{const.}
\end{eqnarray}
In the quantum spin liquid, 
\(\la f^{\dagger}_{i, \a, \ell}f^\pdg_{j, \b, \ell'} \ra \!=\! \d_{\a\b}\d_{\ell, \ell'}\c_{ij}\) with $\chi_{ij} \!\sim\! \mo(1)$.
%for \(i\neq j\), and \(\c_{ii}=1/2\) consistent with half-filling. 
By contrast, deep in a magnetically ordered state, it suffices to consider single-site order with
%\(\la f^{\dagger}_{i, \a, \ell}f^\pdg_{j, \b, \ell'} \ra = \d_{ij}\d_{\ell, \ell'}\left(\sqrt{N} m_i^a T^a_{\a\b}+(1/2)\d_{\a\b}\right)\) 
$\la f^{\dagger}_{i, \a, \ell}f^\pdg_{j, \b, \ell'} \ra \!\sim\! \d_{ij}\d_{\ell, \ell'} \sqrt{N} m_i^a T^a_{\a\b}$,
with $m^a_i m^a_i \sim \mo(1)$. Magnetic ordering corresponds to splitting the $2N$ levels into two sets of $N$ levels and 
occupying the lower $N$ levels, which amounts to switching on Weiss fields which couple to $N$ generators
out of a total of \(4N^2-1\) generators. These \(N\) Weiss fields can be viewed as \(N\) 
copies of a conventional \(SU(2)\) Weiss field, each of which Zeeman splits one pair of the 2N levels.
%corresponding to a \(SU(2)\) ordered state acting on \(N\) different pairs of \(SU(2N)\).
The resulting Hartree-Fock mean-field expectation value of the Hamiltonian thus scales as
%\la \mh_{ij} \ra_{\rm mf} \sim \frac{1}{2} N M  {\mathbb J}_{ij} \left( \frac{M}{2N} \mathbf{m}_i\cdot \mathbf{m}_j- 4|\c_{ij}|^2\right)\,.
$\la \mh_{ij} \ra_{\rm mf} \sim {\mathbb J}_{ij} (-N^2 M |\c_{ij}|^2 + N M^2 \mathbf{m}_i\cdot \mathbf{m}_j)$. 
Scaling the exchange coupling as ${\mathbb J}_{ij} = J/N$ yields
$\la \mh_{ij} \ra_{\rm mf} \sim J_{ij} (-N M |\c_{ij}|^2 + M^2 \mathbf{m}_i\cdot \mathbf{m}_j)$.
 This mean field result, in the 
usual manner, is expected to become exact in the limit of large $M,N \to \infty$ with fixed $\kappa=M/2N$. 
The ratio $\kappa$ plays the role of `spin' in a manner similar to the $Sp(2N)$ framework.
% The ratio of two large numbers \(S\equiv M/2N\) plays the role of \emph{spin-length}, consistent with the fact that a larger spin-length will favor the ordered state. 
We can construct the appropriately defined local moment sum-rule in this large-\((N, M)\) limit which yields
$\kappa(\kappa+1)$;
see Appendix \ref{appendix:largeNM}. \par 
Finally, we briefly comment on the impact of fluctuations beyond mean field theory.
In the usual \(SU(2N)\) extension of fermionic theories sub-leading corrections to the
spin susceptibility from gauge fluctuations appear as ${\cal O}(\frac{1}{N})$ corrections
\cite{Hermele_2004_Stability,Hermele2005_ASL}. In our extension, since there again is a single local constraint, the
gauge field will couple to all the $NM$ degrees of freedom, and we thus expect gauge fluctuation corrections to spin correlations
to be more strongly suppressed as ${\cal O}(\frac{1}{NM})$ or equivalently ${\cal O}(\frac{1}{\kappa N^2})$.
%This will hold true in our extension as well and hence gauge corrections to spin susceptibility will be suppressed by \(\mo(1/N_f)\propto 1/(2NM)\). 
The spin response will however also get contributions from order parameter fluctuations - the leading order corrections will come 
from bubble diagrams within a random phase approximation which will contribute at ${\cal O}(\frac{1}{N})$, 
and hence will be more important than gauge 
fluctuation corrections for $M \gg 1$; see Appendix \ref{appendix:largeNM}. In Fig.~\ref{fig:schematic}, we summarize this
scenario for competing spin liquid and magnetically ordered phases.

\subsection{Phenomenological mean field theory of Heisenberg models}

From the above formulation of the spin model, we see that the effect of $\kappa=M/2N$ is to reweight the energy contribution of the
quantum spin liquid relative to magnetically ordered broken symmetry ground states. This reweighting factor can be recast phenomenologically
by splitting the Hamiltonian $\mh_{ij}$ as a sum of two terms, 
\begin{equation}
    \mh_{ij} = \alpha \mh_{ij} + (1-\alpha) \mh_{ij},
    \label{eq:decouple}
\end{equation} 
and doing a mean
field theory where we decouple the first term in a parton hopping channel (favoring a spin-liquid state), and the second term in a 
magnetic ordering channel assuming a local Weiss field (favoring a symmetry broken state). From the previous subsection, we can
identify the ratio $\alpha/(1-\alpha) \!\equiv\! N/M$, so that $\alpha \!=\! N/(N+M) \!=\! 1/(1+2\kappa)$. Taking $\kappa\!\to\!\infty$
corresponds to $\alpha\!=\!0$ which leads to a classical mean field treatment of the spin model, while the limit $\kappa\!\to\! 0$ 
leads to $\alpha\!=\!1$ which is the pure spin liquid limit. The physical model in this picture $\kappa\!=\! 1/2$ corresponds to $\alpha\!=\!1/2$.
We can however explore tuning $\alpha$ (equivalently $\kappa)$ as a way to explore both magnetically ordered and spin liquid phases
and the quantum phase transition between them.

\subsection{Phenomenological mean field theory of anisotropic spin-exchange models}

If the spin-\(1/2\) Hamiltonian has anisotropies which break the \(SU(2)\) spin-rotation symmetry, there is no obvious large-\((N, M)\) extension.
Nevertheless, motivated by the Heisenberg limit, one can still extend the phenomenological mean-field treatment from the previous subsection
to such anisotropic models.
We modify Eq.\eqref{eq:decouple} to have symmetry-distinct parameters corresponding to different generators, and as well as
incorporating possibly bond-dependence (e.g., first-neighbor versus second-neighbor). 
In other words, an anisotropic spin Hamiltonian on a bond \(\mh_{ij}\) should be decomposed into a hopping channel (favoring a spin-liquid state), and a Weiss channel (favoring an ordered state), with some relative ratio of \(\a_{ij}^{a}:(1-\a_{ij}^{a})\) respectively. One can constrain the
parameters in such an approach, i.e. fix the bond and component dependent parameters \(\a^a_{ij}\), using ground state properties like the phase boundaries from other numerical methods, or known order parameters. Once fixed, we can use these parameters to explore excitations via the RPA corrected spin response
function which can be compared with spin dynamics observed in neutron scattering or resonant inelastic X-ray scattering experiments.

\section{Triangular lattice antiferromagnets}

In the following sections, we will use the methodology introduced in Sec.\ref{methodology} and apply it to a specific spin model. Namely, we look at a \(J_1\)-\(J_2\) XXZ spin model on the triangular lattice which consists of anti-ferromagnetic easy-axis interaction between nearest-neighbor and second-nearest-neighbour spins. The Hamiltonian on a bond has the form
\begin{equation}
    \label{easy plane Hamiltonian}
    \mh_{ij} = \left(S_i^x S_j^x + S_i^y S_j^y + \D S_i^z S_j^z\right)\,,
\end{equation}
where \(\D\geq 1\) represents the easy-axis anisotropy. The full Hamiltonian looks like
\begin{equation}
    \label{easy plane Hamiltonian full}
    \mh = J_1\sum_{\expval{i, j}_1}\mh_{ij} + J_2\sum_{\expval{i,j}_2}\mh_{ij}\,,\:\:\:J_1, J_2>0\,.
\end{equation}
where \(\expval{i, j}_1\) and \(\expval{i, j}_2\) is a shorthand for first-nearest and second-nearest neighbors on the triangular lattice. 
We will explore the impact of varying the anisotropy $\Delta$ away from its Heisenberg point $\Delta\!=\!1$, and the strength of the second
neighbor exchange which can frustrate conventional $120$-degree magnetic order of the Heisenberg model.

\subsection{Spinon mean field ground state}
In this section, we analyze the spin Hamiltonian in Eq.\eqref{easy plane Hamiltonian full} at the mean-field level. We first study the Hamiltonian in the Heisenberg limit  \(\D=1\) using our large-\((N, M)\) formalism and later modify the approach for the anisotropic case \(\D>1\).

\subsubsection{Heisenberg Limit \(\D=1\)}
\begin{figure}[t]
 \centering
 \includegraphics[width=0.48\textwidth]{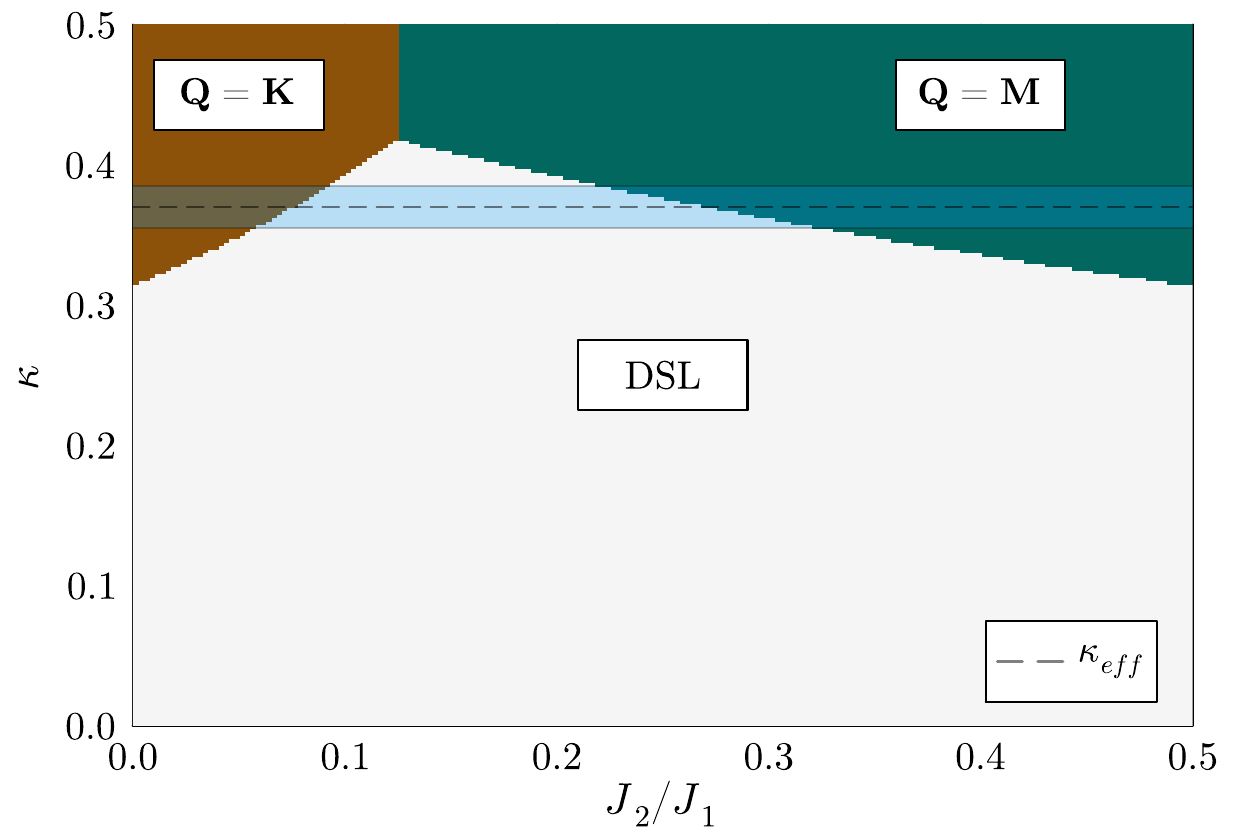}
 \caption{The large-\((N, M)\) phase diagram of the \(SU(2N)\times SU(M)\) generalization of the \(J_1-J_2\) Heisenberg Hamiltonian on the triangular lattice as a function of \(J_2/J_1\) and \(\kappa=M/2N\). At \(\kappa=1/2\), our formalism is equivalent to usual Hartree-Fock mean-field theory which always favors classically ordered states. As the spin \(\kappa\) is decreased, the ordered states transition into disordered spin liquid states, which in this case is a generalization of the \(\pi\)-flux DSL state. Of special interest is that for \(\kappa\approx 0.35-0.37\), the mean-field phase diagram matches qualitatively with many-body results from DMRG \cite{DMRGWhite, DMRG_Sheng, Chernyshev2024} and VMC \cite{J1J2_triangular, J1J2_dynamics}. The \(\bQ=\bM\) order gaps out the system immediately whereas the \(\bQ=\bK\) order is gapless near the phase boundary with the DSL, but gets gapped out eventually.}
\label{fig:large N instabilities}
\end{figure}

We consider a Hamiltonian of the form given in Eq.\eqref{large-N,M Hamiltonian} on nearest and second-nearest neighbors on the triangular lattice in the \(N, M\gg 1\) limit, extending the Heisenberg model. We get a phase diagram tuning the spin \(\kappa=M/2N\), and exchange ratios \(J_2/J_1\). The self-consistent mean-field parameters we allow for in this analysis correspond to a \(SU(2N)\times SU(M)\) symmetric hoppings for the spinons \(f_{i, \a, \ell}\) as \(t_{ij}^{\a\b ; \ell\ell'} = t_{ij}\d^{\a\b}\d_{\ell\ell'}\). We also allow for local Weiss fields corresponding to a \(SU(2N)\) generalization of magnetic orders for each flavor \(\ell\). Any conventional \(SU(2)\) order can be generalized to \(SU(2N)\times SU(M)\) by simply having \(N\) copies of the \(SU(2)\) Weiss fields per flavor \(\ell\) in \(SU(M)\). In principle one can expect more complicated orderings similar to multi-polar ordered state in the \(SU(2N)\) limit. However, for simple Hamiltonians which exhibit collinear and coplanar orderings, any \(SU(2N)\) extensions deep in the ordered phase can be shown to not have lower energies than just \(N\) copies of the \(SU(2)\) order, and hence we can restrict our Weiss fields for simplicity. The resultant phase diagram is shown in Fig.\ref{fig:large N instabilities}. 

We find that the ground state for \(\kappa\lesssim 0.3\) is the generalization of the symmetric \(\pi\)-flux state on the triangular lattice which exhibits two Dirac nodes nested by wave-vectors \(\bQ=\bM_{1/2/3}\) in the Brillouin zone. This state does not break any internal and lattice symmetries. At the mean-field level, the second neighbor exchange interaction does not self-consistently generate any hoppings which may gap out the Dirac nodes. The nearest-neighbor hoppings in this Dirac spin liquid (DSL) can be determined self-consistently in terms of the exchange interactions and the strength is found to 
be \(|t_1|\approx 0.2 N J_1\); see Appendix.\ref{appendix:MFT}. We also find a range of spin \(\kappa \approx 0.35-0.37\) where the phase boundaries match qualitatively with many-body results from DMRG \cite{DMRGWhite, DMRG_Sheng, Chernyshev2024} and VMC \cite{J1J2_triangular, J1J2_dynamics}.\par

\begin{figure}[t]
 \centering
 \includegraphics[width=0.45\textwidth]{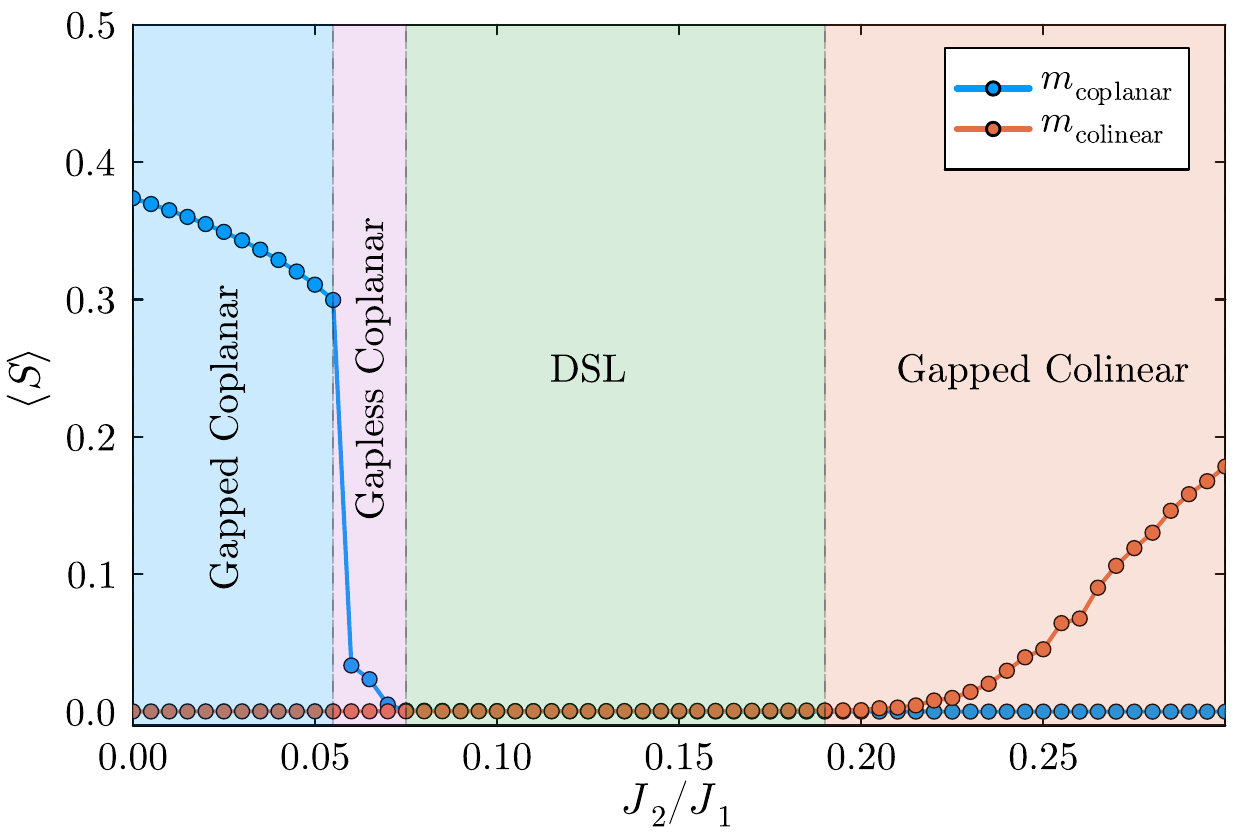}
 \caption{The induced order moments in the \(SU(2N)\times SU(M)\) formalism at a fixed \(\kappa=0.36\) (\(\a\approx 0.6\)). The gapped/gapless notation refers to the mean field parton gap in the spin liquid or ordered phases. For small \(J_2/J_1\leq 0.05\), the ordering is a conventional coplanar ordered state where the fractionalized partons are gapped, and the gauge bosons get confined. Similarly, at large \(J_2/J_1 \geq 0.18\), the system is again a conventional collinear ordered state. Since we consider \(SU(2)\) symmetric spin models, there exist gapless spin-1 Goldstone modes in the conventional orders corresponding to broken symmetries. For intermediate values of \(0.07\leq J_2/J_1\leq 0.18\), the system is in a DSL phase with gapless fractionalized spin-1/2 excitations. Interestingly, there is a small range of \(J_2/J_1\) where coplanar magnetic order
coexists with gapless parton excitations.}
\label{fig:large N order parameters}
\end{figure}

Then we study the order parameters generated at an ``effective spin" of \(\kappa\approx 0.36\) and plot the results in Fig.\ref{fig:large N order parameters}. At \(J_2/J_1=0\), we find an ordered moment of roughly \(\expval{S}\approx 0.35\) (out of a maximum of \(0.5\)), which is in good qualitative agreement with previous results on the nearest-neighbor Heisenberg model on the triangular lattice using a variety of methods. Spin-wave corrections \cite{ChubokovLargeS} and series expansions \cite{SeriesExpansion_Heis} find \(\expval{S}\approx 0.2-0.3\), while ED \cite{HeisED1, HeisED2}, VMC \cite{ArunSupersolid}, and DMRG \cite{DMRGWhite, DMRG_Sheng, Chernyshev2024} find $\la S \ra \approx 0.3$-$0.4$. The effective spin being \(\kappa\approx 0.36\) is equivalent to the phenomenological parameter \(\a\approx 0.6\). We also note that the order parameter will be further reduced due to spin-wave fluctuations.

\subsubsection{Easy-axis anisotropy \(\D>1\)}
Now we introduce an easy-axis anisotropy in the Hamiltonian with \(\D>1\). As mentioned in Sec.\ref{methodology}, we now have to introduce a phenomenological parameters \(\a_{ij}\) which have bond-dependence. Moreover, since the \(SU(2)\) symmetry is broken, generically one can have \(\a_{ij}^{\perp}\neq \a_{ij}^{z}\). For small anisotropies, one can split the full Hamiltonian into a Heisenberg part, and a small deviation from it. Subsequently, the Heisenberg part can be treated similar to the \(SU(2N)\times SU(M)\) formalism with a fixed uniform \(\a\approx 0.6\) as seen in the previous section. The deviation however needs an independent tuning parameter \(\d\a_{ij}\). \par
\begin{figure}[!t]
 \centering
 \includegraphics[width=0.45\textwidth]{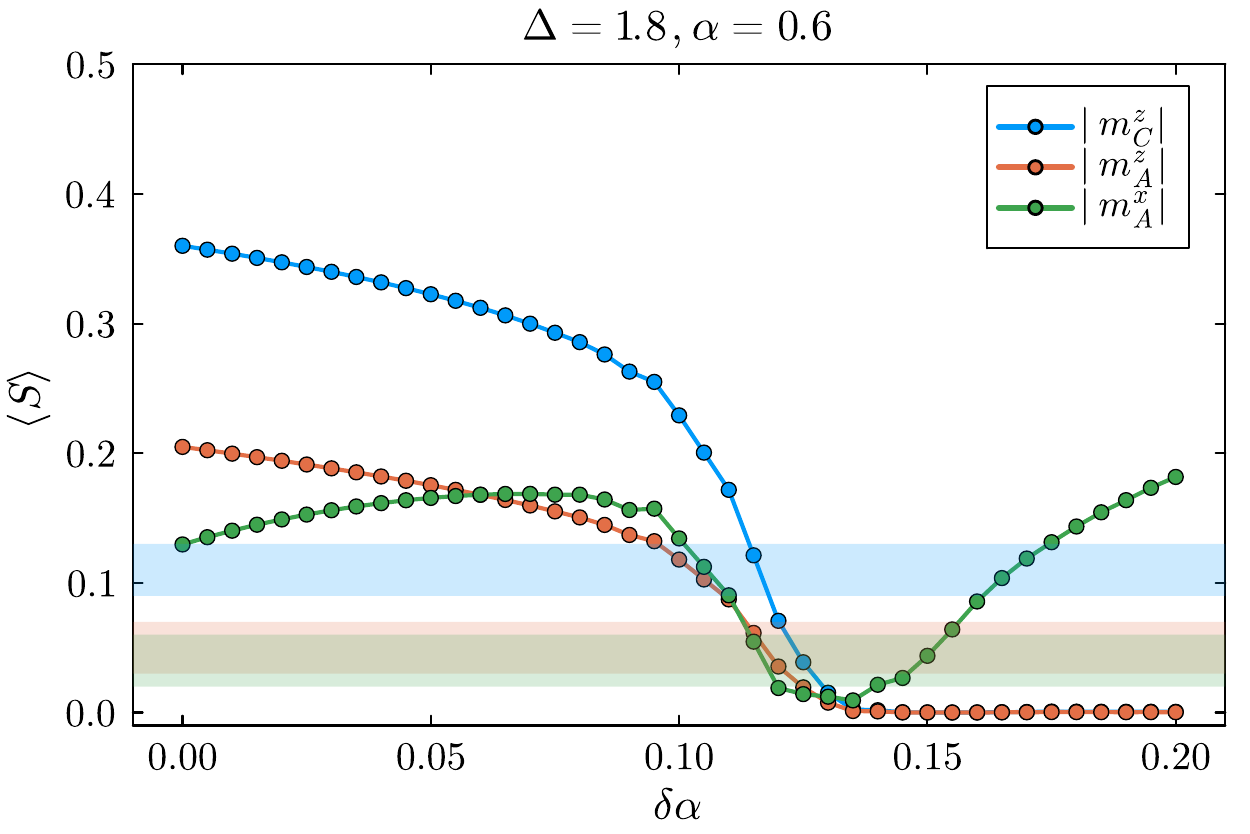}
 \caption{The three independent order parameters of the supersolid order such that the magnetization on the three sublattices of the unit cell look like \(\mathbf{m}_A = (-m_A^x, 0, -m_A^z)\), \(\mathbf{m}_B = (+m_A^x, 0, -m_A^z)\), and \(\mathbf{m}_C = (0, 0, m_C^z)\) at an anisotropy of \(\D=1.8\) relevant for Na$_2$BaCo(PO$_4$)$_2$. The Heisenberg part of the Hamiltonian is treated same as before with \(\a\approx 0.6\), while ``small" deviations due to anisotropy gets decomposed using a phenomenological parameter \(\d\a\). The shaded regions (with corresponding colors) are the experimentally observed values of these moments in 
 Na$_2$BaCo(PO$_4$)$_2$ \cite{NBCOExperiment, SuperSolidGang}.}
 \label{fig:anisotropic order parameter}
\end{figure}

Let us focus on just the nearest neighbor part of the Hamiltonian at a small anisotropy \(\D\). As mentioned above, one needs two parameters \(\a_1^{\perp}\) and \(\a_1^{z}\) which, for small anisotropies, can be recast as \(\a_1^{\perp} = \a - (\D-1)\d\a\) and \(\a_1^z = \a + (\D-1)\d\a\). The classical order that we expect is the so-called \emph{Y-state} which has a three sublattice magnetic unit-cell and consists of both in-plane and out-of plane spin components. This Y-state is also sometimes called a supersolid state in the language of hardcore bosons \cite{SupersolidBosons, HardcoreSupersolid_Ashwin, HardcoreSupersolid_Balents, HardcoreSupersolid_Stefan, HardcoreSupersolid_Massimo}. The out-of-plane spin component of the ordering is like a boson density solid, whereas the in-plane spin component corresponds to the phase of the boson in a superfluid state. In Fig.\ref{fig:anisotropic order parameter}, we plot how the three independent order parameter of a supersolid order develop as we tune \(\d\a\) with a fixed \(\a\approx 0.6\) as motivated from the effective spin \(\kappa\approx 0.36\) in the Heisenberg limit. We find that all three order parameters lie in the experimentally observed regime \cite{NBCOExperiment, SuperSolidGang} for \(\d\a\approx 0.1-0.15\). Furthermore, the effective parton interaction in the phenomenological Hamiltonian Eq.\eqref{eq:decouple} is in good agreement with \cite{SuperSolidGang} where they study a similar approach of residual interactions generating Weiss fields in the Dirac state.

\subsection{Gutzwiller projected wavefunction}
\begin{figure}[!ht]
 \centering
 \includegraphics[width=0.48\textwidth]{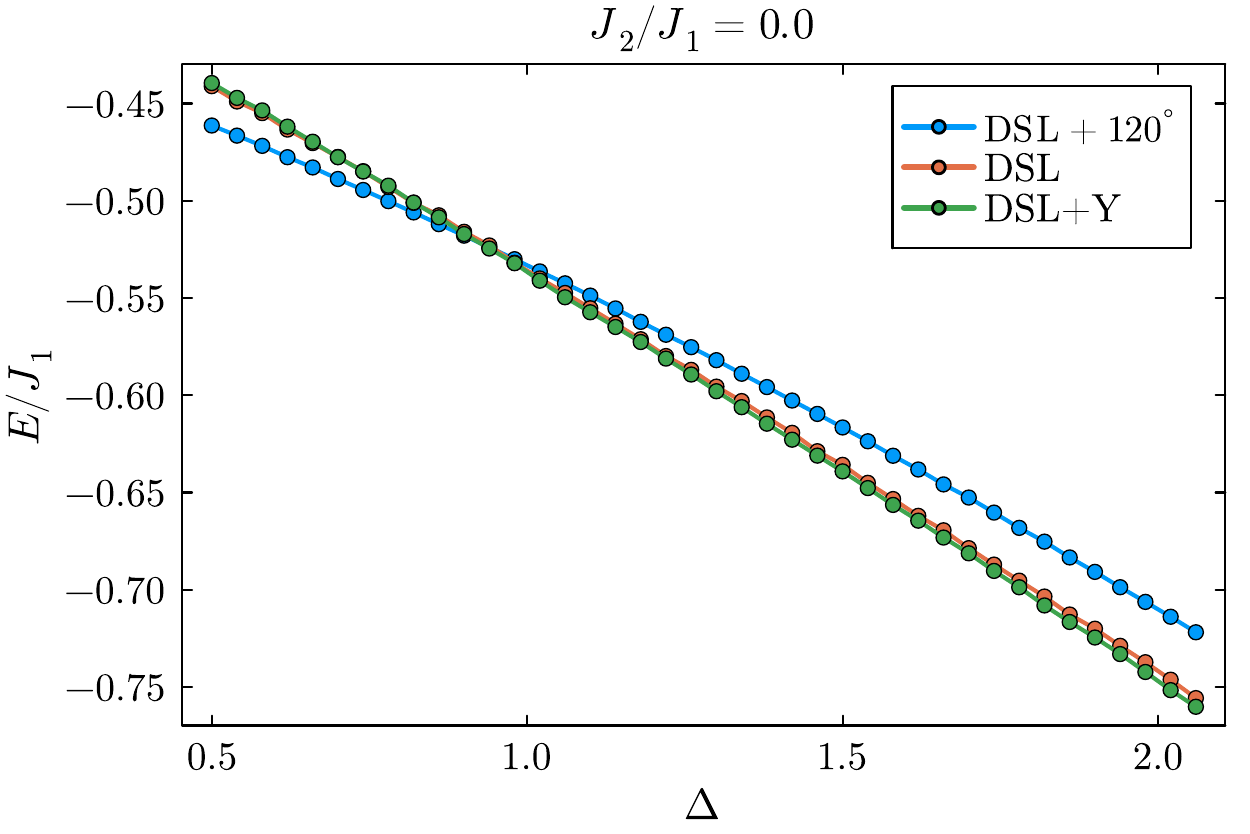}
 \caption{Optimized Gutzwiller projected wavefunction energies for different ansatzes on a finite size system computed using VMC with stochastic optimization of variational parameters including Jastrow factors. The easy-plane Dirac spin liquid (DSL) ansatz has nearest-neighbour hoppings $\pm t_1$ in a \((\pi,0)\) flux state, and Jastrow factors. The states marked ``DSL+order'' incorporate additional Weiss fields as variational parameters which fix the specific type of broken spin symmetry. We find a transition from a coplanar \(120^{\circ}\) ordered state to the supersolid Y-state at the Heisenberg point, \(\D=1\) for a fixed \(J_2/J_1=0\).}
\label{fig:VMC energies Delta}
\end{figure}

Going beyond mean field theory, we also studied the phase diagram of the $J_1$-$J_2$ XXZ model using Gutzwiller projected parton wave-functions. We compute the energy and correlations of the projected wave-function through a Monte Carlo sampling of real space configurations which obey the strict Hilbert space constraint (that the fermionic spinons are at half-filling on each site). There has been previous variational numerics on the easy-axis nearest-neighbour XXZ model \cite{ArunSupersolid}, as well as the Heisenberg limit of the \(J_1-J_2\) model on the triangular lattice \cite{J1J2_triangular, J1J2_dynamics}. In this paper, we extend those numerics by looking at the easy-axis limit of the \(J_1-J_2\) XXZ model on the triangular lattice.\par
From the mean-field theory, we expect to find a supersolid order at small \(J_2/J_1\), a stripe order at large \(J_2/J_1\) with possibly an intermediate spin liquid phase. Keeping that in mind, the variational parameters we allow for in our analysis are independent Weiss fields which can accommodate both the supersolid order and the stripe order, parton hoppings in the \(\pi\)-flux state, as well as long-range isotropic Jastrow factors ; see Appendix \ref{appendix:VMC}.

\begin{figure}[b]
 \centering
  \includegraphics[width=0.48\textwidth]{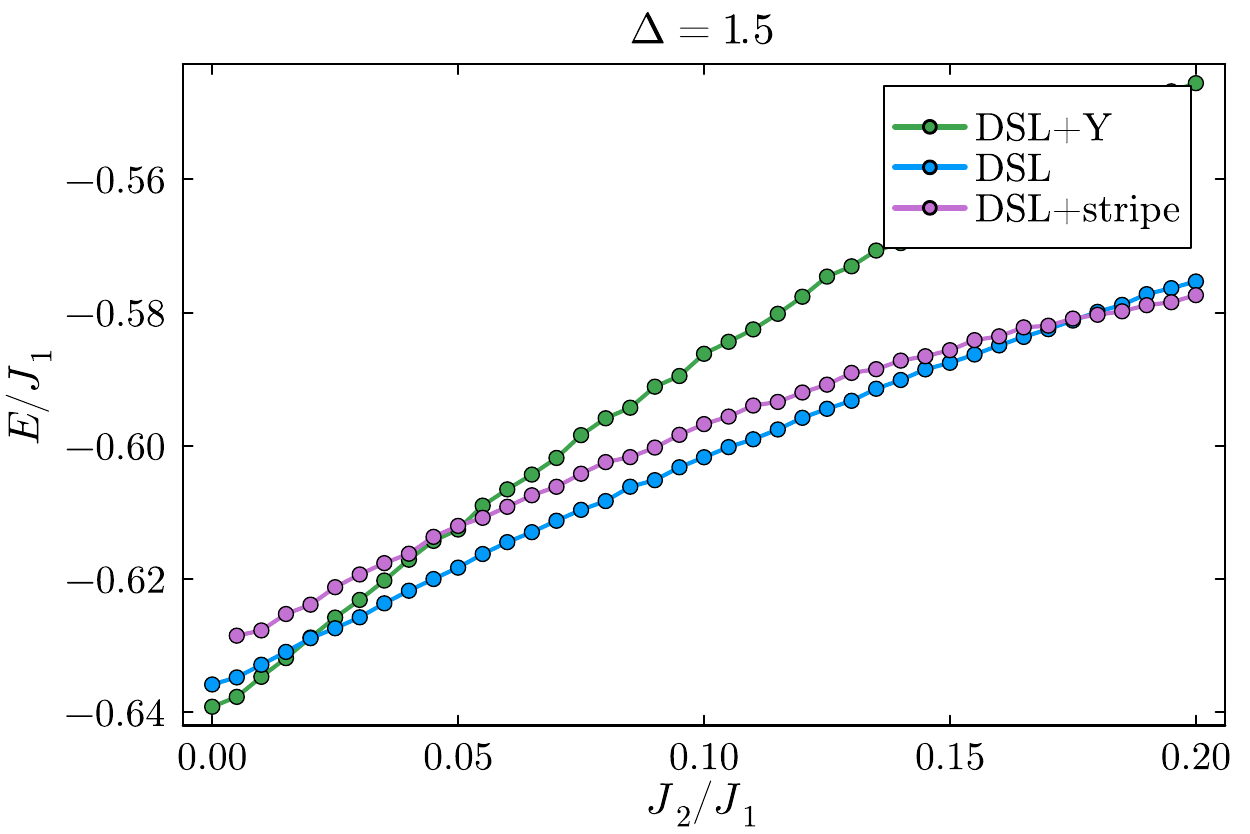}
 \caption{Optimized VMC energies similar to Fig. \ref{fig:VMC energies Delta} but for a fixed anisotropy of \(\D=1,5\). We find that the supersolid order Y-state transitions into the DSL state at a small value of \(J_2/J_1\lesssim 0.04\), followed by a transition into the out-of-plane z-stripe order at \(J_2/J_1\gtrsim 0.16\). These are in good quantitative agreement with DMRG results in \cite{Chernyshev2024}.}
\label{fig:VMC energies J2}
\end{figure}

The results of the optimum energies in the DSL state, as well as DSL+order ansatzes are shown in Fig.\ref{fig:VMC energies Delta} and Fig.\ref{fig:VMC energies J2}. For a fixed \(J_2/J_1=0\), we find the ground state to be the coplanar \(120^{\circ}\) ordered state when the anisotropy is easy-plane, \(0.5 \leq \D < 1\). Meanwhile, when the anisotropy is easy-axis, \(1 < \D \leq 2\), we find the ground state to be the Y-state, with the energetics consistent with previous VMC work \cite{ArunSupersolid}. Throughout, the pure DSL is closeby energetically and for that reason we posit that it can be treated as a parent spin liquid state for these symmetry broken phases \cite{Hermele2005_ASL}. 
Next, for a fixed anisotropy of \(\D=1.5\), we find the supersolid Y-state gives way to the pure DSL as the true ground state at small \(J_2/J_1 \approx 0.03-0.04\). As \(J_2/J_1\) is increased, eventually the pure DSL transitions into the out-of-plane z-stripe order at \(J_2/J_1\approx 0.16-0.17\). These are in good agreement with recent DMRG results\cite{Chernyshev2024}.

\section{Spin dynamics}
In this section, we report the spin dynamics of the mean-field states after incorporating RPA corrections coming from magnon like excitations of the internal Weiss fields. First, we calculate the sublattice-resolved bare mean-field spin susceptibility of the spinons \cite{Wen2002, Wen2015, Iqbal2020, YBK2003, BOcquet2001} using a simple bubble diagram 
\begin{equation}
   \label{chi bubble}
    \c_{0, ij}^{ab}(\bQ, i\Omega)\! =\!-\!\!\sum_{\bk,\;i\omega}  \!\! \text{Tr} [\frac{\s^{a}}{2}\mg_{ij}^{T}(\bk, i\w) \frac{\s^{b}}{2} \mg_{ji}^{T}(\bk+\bQ, i\w+i\Omega)]\,,
\end{equation}
where \(a\,,b\) represent spin-directions, and \(i\,,j\) mark sublattices, and \(\mg\) is the bare spinon Green's function. We note that this bare mean-field susceptibility can be calculated in both the symmetric DSL state, as well as broken symmetry DSL+ordering state. Following this, we can analytically continue Eq.\eqref{chi bubble} into real-frequency as \(\c_{0, ij}^{ab}(\bQ, \W+i\eta)\). Incorporating RPA corrections to the response, we get
\begin{equation}
    \label{RPA bubble}
    \!\! \c_{ij}^{ab}(\bQ, \W) \!=\! \left[(\mathds{1}\!-\!\mj(\bQ)\!\cdot\!\c_0(\bQ, \W))^{-1}\!\cdot\! \c_0(\bQ, \W)\right]_{ij}^{ab}\,,
\end{equation}
where all algebraic operations like \(\mj\cdot \c_0\), or \(\c_0^{-1}\), are shorthands for \emph{matrix operations} in both spin-space, as well as sublattice space. The effective interaction responsible for the RPA scale as \(\mj_{ij}\sim(1-\a_{ij})J_{ij}/\a_{ij} t_{ij}\) where \(t_{ij}\) is the self-consistent Dirac hopping generated \(|t_1|\approx 0.2 J_1 (1+\D)\) for small anisotropies. Lastly, since we want to compare with experimental results, we Fourier transform over actual real-space position, getting rid of the sublattice index, and obtain \(\c_{ab}(\bQ, \W) = \sum_{i, j}\c_{ij}^{ab}(\bQ, \W) e^{-i \bQ\cdot(\br_i-\br_j)}\). The full dynamic spin structure factor at zero-temperature then is simply \(\bS(\bQ, \W) = (1/\pi)(\c_{xx}^{''}(\bQ, \W)+\c_{yy}^{''}(\bQ, \W)+\c_{zz}^{''}(\bQ, \W))\).

Below, we compare the results of this approach with experimental data on a variety of recently studied triangular lattice antiferromagnets. The published experimental data are presented as false color plots of the scattering intensity to emphasize certain features; we have not aimed to reproduce the precise color scales, but instead we focus on highlighting that our approach captures several key features in the data as we discuss below.

\subsection{Heisenberg limit \(\D=1\)}
\begin{figure}[b]
 \centering
    \includegraphics[width=0.5\textwidth]{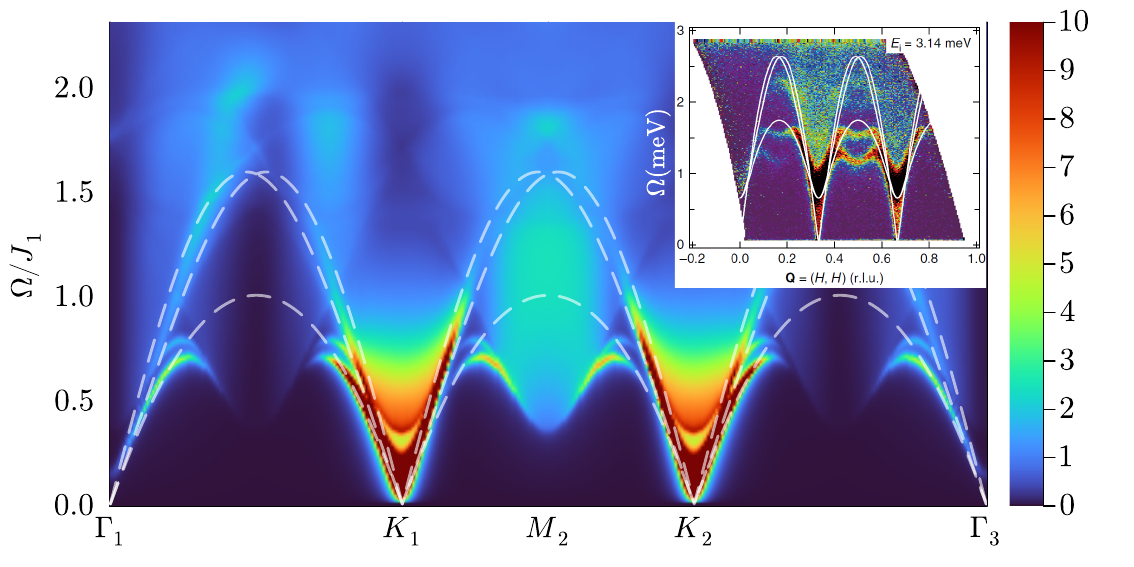}
 \caption{The dynamic spin structure factor \(\bS(\bQ, \W)\) measured along a high-symmetry path in the Brillouin zone and throughout the bandwidth for the nearest neighbor Heisenberg model on the triangular lattice. The dashed white lines mark the spin-waves calculated around the \(120^{\circ}\) coplanar order which have three zero-modes at the \(\bGa\) and \(\bK\) points in the Brillouin zone. The inset shows the experimental results in \bcso\,\cite{bcso_nature} for comparison with \(J_1\approx 1.7\)meV and \(\Delta\approx 0.95\) \cite{bcso_nature}. The mean-field+RPA spin response exhibits sharp modes just like the spin-waves near the \(\bGa\) and \(\bK\) points followed by a weak continuum of spectral weight. However, near the \(\bM\) and \(\bK/2\) points, the spectrum differs from simple spin-waves and in fact a sharp mode comes down in energy, while showing a broad continuum as well.}
\label{fig:Inelastic Neutron plot Heis}
\end{figure}

Fig.\ref{fig:Inelastic Neutron plot Heis} shows the real frequency dynamic spin structure factor $\bS(\bQ, \Omega)$ for momenta along a high-symmetry path and all energies. Near the \(\bGa_1=(0, 0)\) and \(\bGa_3=(2\pi, 2\sqrt{3}\pi)\) points we find clear sharp weakly gapped modes, while at the \(\bK_1=(2\pi/3, 2\pi/\sqrt{3})\) and \(\bK_2=(4\pi/3, 4\pi/\sqrt{3})\) points, we find clear sharp gapless modes matching with magnon bands calculated from linear spin-wave theory around a coplanar order, with a continuum background coming from the spinons. Near the \(\bM_2=(\pi, \sqrt{3}\pi)\) and \(\bK_{1/2}/2\) points, we find a sharp soft mode which is approximately at half the energy of the expected lowest spin wave, along with a broad continuum coming from nesting of the Dirac nodes. The existence of a sharp mode coming down in energy at these momenta is in very good agreement with dynamics calculated through other methods like series expansion \cite{SeriesExpansion_Heis, SeriesExpSpectra, SeriesExpansionJ1J2}, magnon-magnon interaction effects \cite{magnonDecay1, magnonDecay2}, VMC \cite{J1J2_dynamics}, and DMRG \cite{DMRG_dynamics, J1J2_Moore}. Moreover, our results are in good qualitative agreement with INS experiments on \bcso\,\cite{bcso_dynamics, bcso_nature, bcso_prl, bcso_Huang_2022} which has been reported to have a small easy-plane anisotropy of \(\D\approx 0.95\), and with KYbSe$_2$ which is modeled by a Heisenberg spin model with \(J_2/J_1\approx 0.05\) \cite{Scheie2024}.\par 

The advantages of our method are twofold. Firstly, not only do we capture the existence of these sharp magnon bands coming from linear spin wave theory, but also the continuum of spectrum expected from fractionalized excitations. The  broad response at low energies \textit{e.g.} around the \(\bM\) points is natural in the spinon language as coming from nesting of the Dirac nodes, a feature which even magnon-magnon interactions are not able to capture. Secondly, our approach is numerically much less intensive than calculating dynamics in many-body techniques like VMC and DMRG. \par

\subsection{Easy-axis anisotropy \(\D>1\)}

\begin{figure}[!ht]
 \centering
    \includegraphics[width=0.5\textwidth]{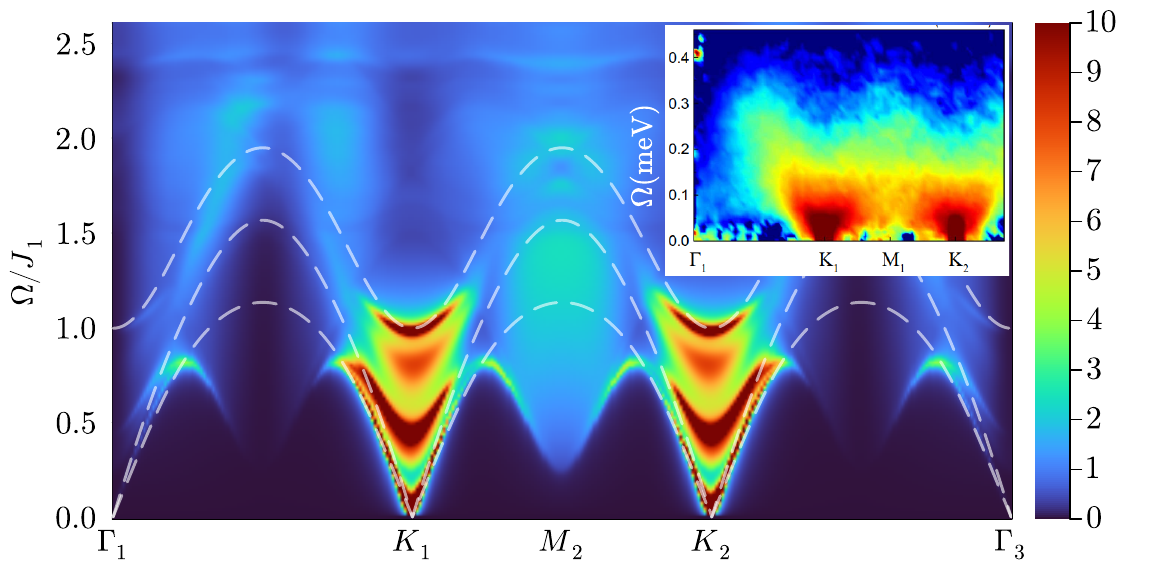}
 \caption{ The dynamic spin structure factor \(\bS(\bQ, \W)\) measured along a high-symmetry path in the Brillouin zone and throughout the bandwidth for the nearest neighbor XXZ model on the triangular lattice with \(\D=1.8\). The dashed white lines mark the spin-waves calculated around the supersolid order which have two zero-modes at the \(\bGa\) and \(\bK\) points in the Brillouin zone. Note that one of these zero modes comes from an accidental degeneracy and has nothing to do with the symmetries of the Hamiltonian. The inset shows the experimental results in \ncpo\,\cite{sheng2024continuumspinexcitationsordered} with \(J_1\approx 0.1\)meV and \(\Delta\approx 1.8\) \cite{NBCOExperiment}. The mean-field+RPA susceptibility again exhibits sharp modes just like the spin-waves near the \(\bGa\) and \(\bK\) points with the multiple sharp modes at the \(\bK\) points being separated with a strong continuum. There still exists a sharp mode at the \(\bM\) and \(\bK/2\), along with a gapped continuum coming from the Dirac nodes nesting. }
\label{fig:Inelastic Neutron plots D=1.8}
\end{figure}

In this section, we again calculate the mean-field+RPA spin response but now in the presence of an anisotropy \(\D>1\). We plot the imaginary part of the in-plane susceptibility at a small anisotropy of \(\D=1.8\) relevant for Na$_2$BaCo(PO$_4$)$_2$ in Fig.\ref{fig:Inelastic Neutron plots D=1.8}. We again find clear sharp modes matching with magnon bands calculated from linear spin-wave theory around a supersolid order near the \(\bGa\) and \(\bK\) points. The anisotropy pushes one of the spin-waves higher in energy \(\propto \D\) which is also seen in our RPA plot, with a strong continuum between the low-energy and high-energy modes. There still exists a soft sharp mode at the \(\bM\) and \(\bK/2\) which is lower in energy than expected from spin-waves, along with a gapped continuum coming from the Dirac nodes nesting. This is again in good qualitative agreement with experimental results observed for Na$_2$BaCo(PO$_4$)$_2$ \cite{NBCOExperiment, sheng2024continuumspinexcitationsordered, SuperSolidGang} where similar features have been reported. \par

\begin{figure}[b]
 \centering
    \includegraphics[width=0.5\textwidth]{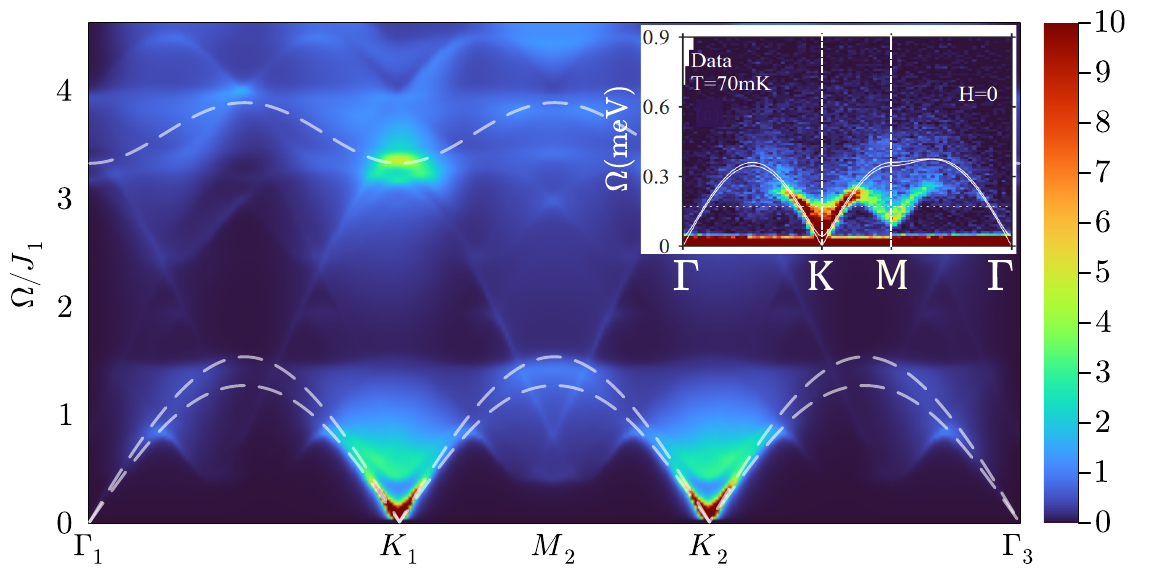}
 \caption{ The dynamic spin structure factor \(\bS(\bQ, \W)\) measured along a high-symmetry path in the Brillouin zone and throughout the bandwidth for the nearest neighbor XXZ model on the triangular lattice with \(\D=10\). The dashed white lines mark the spin-waves calculated around the supersolid order which have two zero-modes at the \(\bGa\) and \(\bK\) points in the Brillouin zone. The inset shows the experimental results in \kcso\,\cite{Zhu_2024} with \(J_1\approx 0.3\)meV and \(\Delta\approx 12\) \cite{chen2024phasediagramspectroscopicsignatures}. Note that the energy scales are different in the theory plots where we show the full spectrum for completeness. Furthermore, for consistency we stick with the same high-symmetry path as in Fig.~\ref{fig:Inelastic Neutron plot Heis} and Fig.~\ref{fig:Inelastic Neutron plots D=1.8}, which is different than the reported experimental plot. The mean-field+RPA susceptibility again exhibits sharp modes just like the spin-waves near the \(\bGa\) and \(\bK\) points with the multiple sharp modes at the \(\bK\) points being separated with a strong continuum. At the \(\bM\) and \(\bK/2\), there is a weak gapped continuum coming from the Dirac nodes nesting. }
\label{fig:Inelastic Neutron plots D=10}
\end{figure}

\begin{figure}[!ht]
 \centering
    \includegraphics[width=0.425\textwidth]{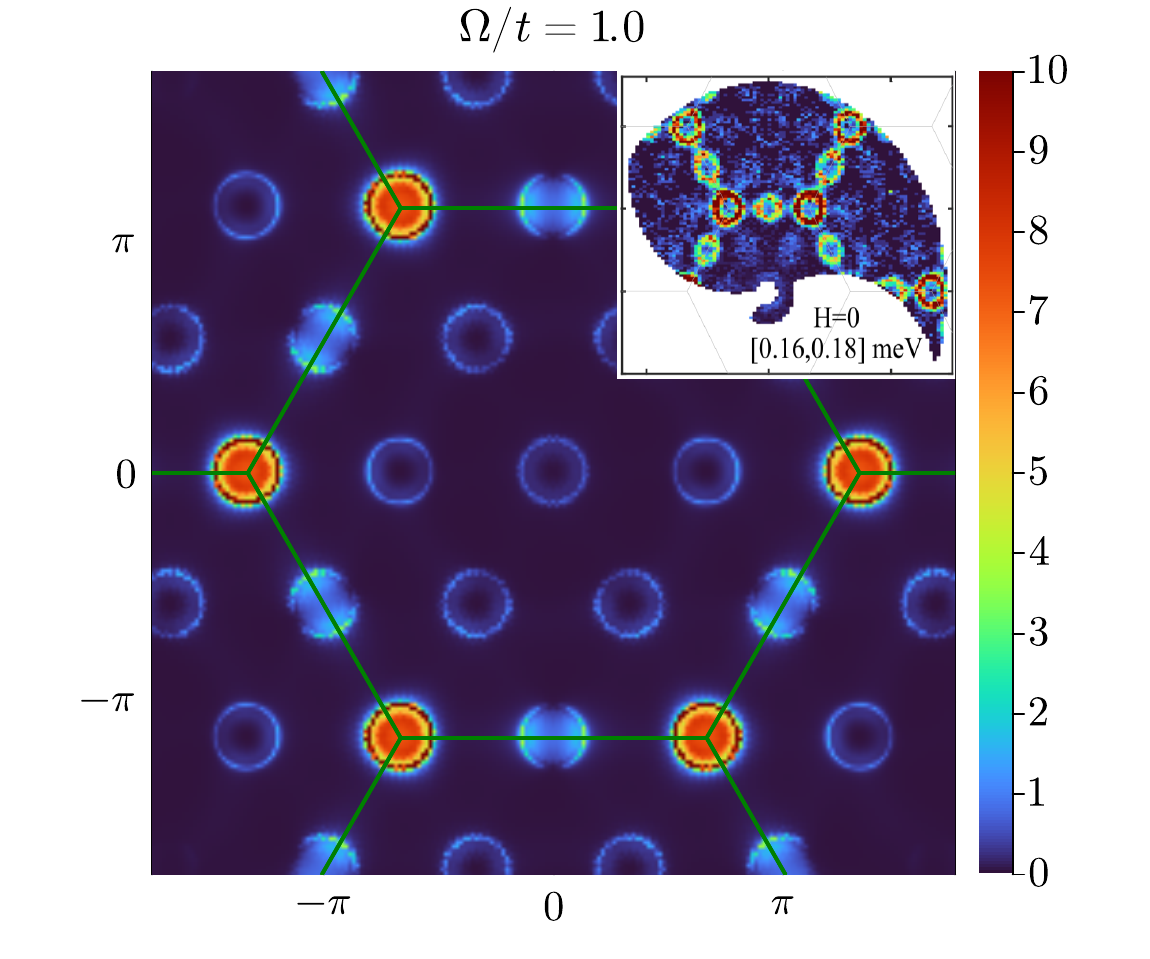}
 \caption{ The dynamic spin structure factor \(\bS(\bQ, \W)\) measured on the entire Brillouin zone at a fixed energy \(\W=t\approx 0.4J_1\) for the nearest neighbor XXZ model on the triangular lattice with \(\D=1.8\). The response shows strong spectral weight peaks at the \(\bK\) points and the \(\bM\) points, as well as weaker features at the \(\bGa\) and \(\bK/2\) points. The low energy behavior of the susceptibility should not be affected much by increasing the anisotropy. Therefore, this can be compared with the inset plot which shows the experimental result for K$_2$Co(SeO$_3$)$_2$ \cite{Zhu_2024}, with which we find good qualitative agreement. Note that K$_2$Co(SeO$_3$)$_2$ has \(J_1\approx 0.3\)meV and \(\Delta\approx 12\) \cite{chen2024phasediagramspectroscopicsignatures}, for which \(\W=t\approx 0.7J_1\). }
\label{fig:Full BZ Neutron plots D=1.8}
\end{figure}

We expect from simple spin-wave theory that increasing the anisotropy \(\D\) just pushes up the high-energy spin wave which is out-of-plane and hence its energy scales with \(\D\). We confirm this in our mean-field+RPA approach by repeating the analysis in a large \(\D=10\) regime which is relevant for materials like K$_2$Co(SeO$_3$)$_2$ \cite{Zhu_2024, chen2024phasediagramspectroscopicsignatures} and Rb$_2$Co(SeO$_3$)$_2$ \cite{PhysRevMaterials.4.084406} as plotted in Fig.\ref{fig:Inelastic Neutron plots D=10}. We also plot the susceptibility over the entire Brillouin zone at a fixed low energy of \(\W/t=1\) in Fig.\ref{fig:Full BZ Neutron plots D=1.8} which we expect should be fairly independent of the anisotropy. We find this also to be in good qualitative agreement with experiments on K$_2$Co(SeO$_3$)$_2$ \cite{Zhu_2024, chen2024phasediagramspectroscopicsignatures}.

\section{Discussion}
In this paper, we have shown that a modified extension of the usual large-\(N\) \(SU(2N)\) formalism for partons to a distinct $SU(2N)\times SU(M)$
version, which incorporates
an additional `spin' length $\kappa=M/2N$, is useful in treating gapless spin liquids and magnetically ordered phases democratically while suppressing gauge fluctuations. We have determined the phase diagram for the antiferromagnetic $J_1$-$J_2$ Heisenberg model on the triangular lattice in the 
large-\((N, M)\) limit. The ground state phase diagram and order parameters from the mean field theory match with previously known numerical
results for \(\kappa\!\approx\!0.36\). We phenomenologically adapt this approach to the \(J_1\!-\! J_2\) XXZ model on the triangular lattice which
includes easy-axis anisotropy. In this approach, we split the spin Hamiltonian (written in terms of fermionic partons) into two parts phenomenologically and decompose each using different HS fields: local dynamical Weiss fields, as well as spinon hopping fields assumed to be condensed around their mean-field value. We supplement the mean-field results with Gutzwiller projected wavefunctions studied using VMC where we find the phase diagram of the XXZ model as we tune \(J_2/J_1\) and find good qualitative agreement with recent DMRG result \cite{DMRG_DSL, Chernyshev2024}. For small \(J_2/J_1\), we
find that the ground state is a DSL with
weak supersolid ordering which gaps the Dirac nodes and is thus expected to be a conventional ordered phase. 
With increasing $J_2/J_1$, this supersolid gives way to a pure DSL at intermediate frustration. Finally, for larger $J_2/J_1$, the DSL develops additional weak stripe order which again gaps out the Dirac nodes.\par 
Next we turn to the dynamics of the spin model where we calculate dynamic spin susceptibility beyond mean-field level by including RPA corrections coming from fluctuations of the internal Weiss fields. This is justified for $N,M \!\gg\! 1$ since RPA corrections are the leading order corrections to 
spin-correlations, as compared to gauge fluctuations which are sub-leading. We find the spin-susceptibility in the Heisenberg limit matches extremely well with previous many-body dynamics from VMC\cite{J1J2_dynamics}, DMRG \cite{DMRG_dynamics, J1J2_Moore}, and analytical results from series expansion \cite{SeriesExpansion_Heis, SeriesExpSpectra} and magnon-magnon interactions \cite{magnonDecay1, magnonDecay2}. Lastly, motivated by the Heisenberg limit, we calculate the dynamics for the anisotropic XXZ model and find very good agreement with INS experimental observations on materials like \bcso\,\cite{bcso_dynamics, bcso_nature, bcso_prl, bcso_Huang_2022}, \ncpo\, \cite{NBCOExperiment, sheng2024continuumspinexcitationsordered, SuperSolidGang}, \kcso\, \cite{Zhu_2024, chen2024phasediagramspectroscopicsignatures}, and \rcso\,.

For future directions, it would be interesting to develop a diagrammatic route to our methodology which can be extended to include parton interaction
diagrams beyond RPA and gauge fluctuation corrections, both of which might perhaps lead to improved quantitative agreement with experimental results. There exists a family of diagrams coming from corrections to the parton propagator which affects the dynamics at leading order compared to gauge fluctuations. However treating those will involve solving self-consistent Dyson equations and we leave that for a future work. Furthermore, it would be useful if a more formal extension can be derived for anisotropic Hamiltonians just like the \(SU(2)\) symmetric Heisenberg Hamiltonians. We note that this is a general problem faced even in the usual \(SU(2N)\) and \(Sp(2N)\) formalisms and is an open question. \par 
It would also be interesting to study and connect our current results to the low energy theory for the DSL which would consist of \(N_f=4\) flavour fermions coupled to a \(U(1)\) gauge theory \cite{Hermele_2004_Stability,Hermele2005_ASL}. Specifically, our approach treats the DSL to ordering transition as being induced by residual four-fermion interactions which have the same form as the original spin Hamiltonian. One needs to understand, however, the role that monopoles play in these possibly weakly ordered Dirac spin-liquids, and how they compete with four-fermion interaction induced symmetry breaking \cite{DiracMonopoles_PRX2020, Song2019}. Lastly, previous \(Sp(2N)\) models have been utilized in doped regimes to understand charge orderings and superconductivity \cite{SuperconductivitySp2N}. It would be of interest to understand if the same can be done in our \(SU(2N)\times SU(M)\) model as well.

Regarding prospective materials where this approach is relevant, there is an abundance of candidates which order at \(T\ll \theta_{CW}\), but exhibit dynamics which are not completely characterized by simple LSWT. There has been a series of work on triangular lattice antiferromagnet {\ybmg} where a similar RPA enhanced mean-field spinon fermi surface state has been successful in describing inelastic neutron scattering experiments \cite{triangleAFM2016, triangleAFM2017, triangleAFM_RPA}. Further, recently, there has been a surge in DSL candidate materials on the triangular lattice such as \ybzn \cite{xu2023realizationu1diracquantum}, and {\cmao} \cite{CMAO}. The INS spectra of {\ybzn} appears to have a very flat continuum like response which does not naively seem compatible with a simple \(\pi\)-flux state on the triangular lattice. Although DMRG recreates some features of the spectra, the underlying microscopic picture is not very clear. It would hence be interesting to understand the claims about DSL in these materials using our approach as well. Investigating these theoretical extensions and candidate materials would improve our understanding of the interplay between spinons, magnons, and gauge fluctuations in the dynamics of frustrated quantum magnets.

\acknowledgments
We acknowledge support from the Natural Sciences and Engineering Research Council (NSERC) of Canada. All numerical computations were performed on the Niagara supercomputer at the SciNet HPC Consortium and the Digital Research Alliance of Canada. The Julia codes for tight-binding analysis and mean-field simulations are available online at \href{https://github.com/Anjishnubose/TightBindingToolkit.jl}{TightBindingToolkit.jl} and \href{https://github.com/Anjishnubose/MeanFieldToolkit.jl}{MeanFieldToolkit.jl} respectively. The bare susceptibility results were done in Julia as well as using TRIQS \cite{Parcollet_2015}

\bibliography{ref}

\appendix 
\onecolumngrid
\clearpage
  \section{\(SU(2N)\times SU(M)\) Partons} \label{appendix:largeNM}
We start with the usual \(SU(2)\) fermionic partons where a local Spin operator \(S_i^a\) is written in terms of fermions as
\begin{equation}
    \label{spin from partons}
    S_i^a = \frac{1}{2}f^{\dagger}_{i, \a}\s^a_{\a\b}f_{i, \b}\,,\:\:\:f^{\dagger}_{i, \a}f_{i, \a}=1 \forall i\,,
\end{equation}
where \(i\) represents a lattice site, \(\a, \b\) are \(SU(2)\) indices in the fundamental representation with \(SU(2)\) generators \(T^a = \s^a/2\) for \(a \in \{1, 2, 3\}\). A generic spin-spin \(SU(2)\) invariant Hamiltonian has the form
\begin{equation}
    \label{generic sym ham}
    \begin{split}
       \mh_{ij} & = J_{ij} S_i^a\cdot S_j^a\\
        & = J_{ij}\left\{\frac{1}{2}f^{\dagger}_{i, \a}f_{i, \b}f^{\dagger}_{j, \b}f_{j, \a} - \frac{1}{4}f^{\dagger}_{i, \a}f_{i, \a}f^{\dagger}_{j, \b}f_{j, \b}\right\}\,,
    \end{split}
\end{equation}
where we have used the completeness relation for \(SU(2)\) generators in the second line. Note that the second quartic fermion term is just a constant since the fillings of the partons are set to 1 exactly (at the operator level itself).\par

We then enlarge the internal symmetry group to \(SU(2N)\) introduce a dummy \(SU(M)\) index on the partons \(f_{i, \a, \ell}\) where \(i\) marks a lattice site, \(\a\in\{1, 2, ..., 2N\}\), and \(\ell\in\{1, 2, ..., M\}\). The Hamiltonian in Eq.\eqref{large-N,M Hamiltonian} is a possible and natural extension of the usual \(SU(2)\) Heisenberg Hamiltonian which is invariant under \(SU(2N)\times SU(M)\). Now the fermion number constraint is still that of half-filling. Hence we have a single \(U(1)\) gauge field coupled to \(2NM\gg1\) flavors of fermions and therefore gauge fluctuations are obviously suppressed. Furthermore, we have already argued in the main text why at the mean-field level such a Hamiltonian favors both hopping and on-site ordering for the partons, depending on the ratio of two large numbers \(\kappa(N, M)\equiv M/2N\). This ratio will turn out to play the role of the \emph{spin-length}, consistent with the interpretation that a larger spin-length favors magnetic ordering and vice-versa. This interpretation can be formalized by looking at spin-sum rules.\par

Since there are infinitely many operators symmetric under \(SU(2N)\times SU(M)\) which reduce to the usual spin-\(1/2\) \(SU(2)\) operators in the \(N=1,M=1\) case, we have a unique and weird freedom on our hands. We can choose the Hamiltonian to have a different form (based on what states we want to be competitive) than the spin-spin correlations in the structure factor, with the constraint that they match in the case of spin-\(1/2\). There are some further restrictions that we should sensibly impose on a structure factor operator \(O_{ij}(N, M)\) :
\begin{itemize}
    \item \(O_{ij}(N=1, M=1) = S_i^a\cdot S_j^a\) : The operator should reduce to the spin-\(1/2\) spin-spin Heisenberg correlation \emph{exactly} for \(N=M=1\).
    \item \(O_{ii}(N, M) = \kappa(\kappa+1)\) for some \(\kappa(N, M)\sim \mo(1)\) : The operator should reproduce the usual spin-sum rules irrespective of \(N, M\). Furthermore, \(\kappa(N=1, M=1)=1/2\).
    \item (Optional) The operator should have contributions from both the disordered phase and ordered phase at the same order in \((N, M)\).
\end{itemize}
Lets look at the sum rule and try to understand how we can get an operator like that. We get a hint from the ordered state energy \(\propto M^2 \mathbf{m_i}\cdot\mathbf{m_j}\). Intuitively, tuning \(N\) favors the spin liquid and tuning \(M\) favors the ordered state. Hence as mentioned earlier, \(\kappa=M/2N\) is roughly playing the role of the \emph{spin-length}. With a little bit of trial and error we can come up with an operator which looks like
\begin{equation}
    \label{large N,M SSF}
    O_{ij} = \frac{1}{N^3}\left\{\frac{1}{2}f^{\dagger}_{i, \a, \ell}f_{i, \b, \ell}f^{\dagger}_{j, \b, \ell'}f_{j, \a, \ell'}+\frac{1}{2}f^{\dagger}_{i, \a, \ell}f_{i, \a, \ell'}f^{\dagger}_{j, \b, \ell'}f_{j, \b, \ell}-\frac{1}{4N}\left(1+\frac{1}{\kappa}\right)f^{\dagger}_{i, \a, \ell}f_{i, \a, \ell}f^{\dagger}_{j, \b, \ell'}f_{j, \b, \ell'}\right\} \,.
\end{equation}
Since the fermion filling is always fixed at half-filling \emph{exactly}, we have that \(f^{\dagger}_{i, \a, \ell}f_{i, \a, \ell} = NM = 2N^2 \kappa\) as an operator identity. Hence the operator in Eq.\eqref{large N,M SSF} can be simplified to
\begin{equation}
    \label{large N,M SSF simplified}
    \mathbf{S}_i\cdot \mathbf{S}_j \xrightarrow{N, M \gg 1} O_{ij} = \frac{1}{N^3}\left\{\frac{1}{2}f^{\dagger}_{i, \a, \ell}f_{i, \b, \ell}f^{\dagger}_{j, \b, \ell'}f_{j, \a, \ell'}+\frac{1}{2}f^{\dagger}_{i, \a, \ell}f_{i, \a, \ell'}f^{\dagger}_{j, \b, \ell'}f_{j, \b, \ell}\right\} -\kappa(\kappa+1)\,.
\end{equation}
Let us now carefully analyze why this operator satisfies all of the aforementioned requirements. Some properties of these operators that will prove useful are as follows
\begin{itemize}
    \item On-site sum rules for the operators in Eq.\eqref{large N,M SSF simplified} \emph{cannot} be found individually without additional operator constraints on the partons. However, we can relate the two as follows
    \begin{equation}
        \label{sum rules individual}
        \begin{split}
            f^{\dagger}_{i, \a, \ell}f_{i, \a, \ell'}f^{\dagger}_{i, \b, \ell'}f_{i, \b, \ell} & = f^{\dagger}_{i, \a, \ell}f_{i, \a, \ell'} \left(\d_{\ell,\ell'}\d_{\b\b}-f_{i, \b, \ell}f^{\dagger}_{i, \b, \ell'}\right)\\
            & = 2N^2M + f^{\dagger}_{i, \a, \ell}f_{i, \b, \ell}f_{i, \a, \ell'}f^{\dagger}_{i, \b, \ell'}\\
            & = 2N^2M + f^{\dagger}_{i, \a, \ell}f_{i, \b, \ell}\left(\d_{\ell',\ell'}\d_{\a\b}-f^{\dagger}_{i, \b, \ell'}f_{i, \a, \ell'}\right)\\
            & = 2N^2M+NM^2-f^{\dagger}_{i, \a, \ell}f_{i, \b, \ell}f^{\dagger}_{j, \b, \ell'}f_{j, \a, \ell'}\,.
        \end{split}
    \end{equation}
    Hence, \(O_{ii} = (1/2N^3)(2N^2M+NM^2)-\kappa(\kappa+1) = \kappa(\kappa+1)\) with \(\kappa(N, M)=M/2N\) which satisfies the sum rule condition \textit{s.t.} \(\kappa(N=1, M=1)=1/2\). Connecting with our \(\a, 1-\a\) decomposition, we find that \(\a\sim 1/(1+2\kappa)\approx 1/2\kappa\) for large \emph{spin-length} akin to a spin-wave expansion.
    \item For the case of \(N=1, M=1\), it turns out that \(f^{\dagger}_{i, \a, \ell}f_{i, \a, \ell'}f^{\dagger}_{j, \b, \ell'}f_{j, \b, \ell}=1\). Therefore, \(O_{ij} \xrightarrow{N=M=1} (1/2)f^{\dagger}_{i, \a, \ell}f_{i, \b, \ell}f^{\dagger}_{j, \b, \ell'}f_{j, \a, \ell'}-(1/4)\) which matches the usual \(SU(2)\) spin-spin Heisenberg correlation.
\end{itemize}

\section{Path Integral}
We first start with the usual \(SU(2)\) symmetric Hamiltonian
\begin{equation}
    \label{full ham}
    \mh = \frac{1}{2}\sum_{i, j}J_{ij}f^{\dagger}_{i, \a}f_{i, \b}f^{\dagger}_{j, \b}f_{j, \a}
\end{equation}
such that the total action looks like (replacing \(f^{\dagger}\) with \(\bar{f}\) in the path integral)
\begin{equation}
    \label{full action}
    \ms[\bar{f}, f, a_0] = \int d\t \left(\sum_{i}\bar{f}_{i, \a}(\partial_{\t}-a_i^0)f_{i, \a} + \mh(\t)\right)\,.
\end{equation}
Now, we are going to decompose the four fermion interaction in the Hamiltonian democratically in the ordered and disordered channel, akin to usual Hartree-Fock. In the disordered channel we need Hubbard-Stratonovich fields which live on the bonds, \(w_{ij}(\t)\), the condensate of which give rise to parton hoppings at the mean-field level. Now, there are two kinds of fluctuations of \(w_{ij} = |w_{ij}|e^{-i a_{ij}}\). The strength modulation must be gapped \cite{QSLgauge_XGWenBook}, and hence can be safely ignored in the low energy physics. The only fluctuations to keep track of are the gauge fluctuations \(a_{\m}\). Now assuming that the mean-field disordered state is a Dirac state such as the \(\pi\)-flux state, one can linearize the Hamiltonian and go to the continuum limit \cite{Hermele_2004_Stability,Hermele2005_ASL}. In this limit, the low energy theory looks like QED-3 with \(N_f=4\) flavors of fermions
\begin{equation}
    \label{QED3}
    \ms = \int d^3x\:\left[ \bar{\psi}_{\a}(x)(i\slashed{\partial}-\slashed{a})\psi_{\a}(x)+\frac{1}{4g^2}F^2\right]\,,
\end{equation}
where \(F^2\) is a Maxwell term generated by integrating out the high-energy fermion modes above some RG scale. Importantly, the Maxwell term is generated at leading order by a fermion bubble diagram, and hence is \(\propto N_f\). Hence, in our extension of \(SU(2N)\times SU(M)\), we will get the continuum action in the disordered phase as
\begin{equation}
    \label{QED3 large NM}
    \ms = \int d^3x\:\left[ \bar{\psi}_{\a, \ell}(x)(i\slashed{\partial}-\slashed{a})\psi_{\a, \ell}(x)+\frac{2NM}{4g^2}F^2\right]\,.
\end{equation}
Note that if we naively look at the scaling of the Hamiltonian in Eq.\eqref{QED3 large NM}, it scales as \(\mo(NM)\). Remember that in the \(N, M \gg 1\) limit Hamiltonian, the disordered state had energy scaling as \(\mo(N^2M)\) and hence the action above is associated with a rescaled exchange \(\mathbb{J}=J/N\). We use this fact in the ordered channel to get
\begin{equation}
    \label{magnetic HS}
    \exp\left(-\frac{1}{N}\mh_{ij}\right) = \int \prod_{a=1}^{4N^2-1} d h_i^a dh_j^a \exp\left(c_1 h_i^a J^{-1}_{ij} h_j^a - c_2 \bar{f}_{i, \a, \ell}T^{a}_{\a\b}f_{i, \b, \ell}h_j^a - c_2 h_i^a\bar{f}_{j, \a, \ell}T^{a}_{\a\b}f_{j, \b, \ell}\right)\,,\:\:\: \frac{c_2^2}{c_1}=\frac{1}{2N}\,,
\end{equation}
where \(h_i^a(\t) = \sum_{\expval{j}}J_{ij} m_j^a\) can be interpreted as the dynamic internal Weiss field generated on site \(i\). We have an overall freedom in choosing the coefficients \(c_1\) and \(c_2\) in Eq.\eqref{magnetic HS}, however again these can be fixed by simple scaling arguments. We want the expectation value of both of the terms in Eq.\eqref{magnetic HS} to come from the ordered state part of \((1/N)\mh\) and hence scale as \(\mo(M^2)\). With the interpretation of \(h_i^a\) as being \(N\) copies of internal \(SU(2)\) Weiss fields, we know already that \(\expval{h_i^a h_j^a}\sim \mo(N)\) and hence we can fix \(c_1=M^2/2N\) and \(c_2=M/2N\). This is consistent since \(\expval{h_i^a \bar{f}_{i, \a, \ell}T^{a}_{\a\b}f_{i, \b, \ell}} \sim \mo(NM)\). Therefore, combining everything, the full \(SU(2N)\times SU(M)\) low energy theory will look like (switching to momentum space)
\begin{equation}
    \label{action large NM}
    \begin{split}
        \ms = &\int d^3k\:\left[ \bar{\psi}_{\a, \ell}(k)\slashed{k}\psi_{\a, \ell}(k)+\frac{2NM}{4g^2}F^2+\frac{M^2}{2N}h^a(k)\mj^{-1}(k)h^a(-k)\right]\\
        +&\int d^3k\:d^3q\:\left[\frac{M}{2N}h^a(q)\bar{\psi}_{\a, \ell}(k+q)T^a_{\a\b}\psi_{\b, \ell}(k) + \bar{\psi}_{\a, \ell}(k+q)\slashed{a}(q)\psi_{\a, \ell}(k)\right]\,,
    \end{split}
\end{equation}
where \(\mj(\bk)\) is the Weiss field propagator dependent on the exchange interactions on the lattice. We can rescale the gauge field \(a\) and the Weiss field \(h\) and rewrite the action as
\begin{equation}
    \label{action large NM 2}
    \begin{split}
        \ms = &\int d^3k\:\left[ \bar{\psi}_{\a, \ell}(k)\slashed{k}\psi_{\a, \ell}(k)+\frac{1}{4}\tilde{F}^2+\tilde{h}^a(k)\mj^{-1}(k)\tilde{h}^a(-k)\right]\\
        +&\int d^3k\:d^3q\:\left[\frac{1}{\sqrt{2N}}\tilde{h}^a(q)\bar{\psi}_{\a, \ell}(k+q)T^a_{\a\b}\psi_{\b, \ell}(k) + \frac{g}{\sqrt{2NM}}\bar{\psi}_{\a, \ell}(k+q)\slashed{\tilde{a}}(q)\psi_{\a, \ell}(k)\right]\,.
    \end{split}
\end{equation}
Therefore we find that the \(N_f=2NM\) flavors of fermions at low energy are interacting with \(N_b=N^2-1\) flavors of bosons \(\tilde{h}^a\) with a coupling of \(g_h = 1/\sqrt{2N}\), and with a \(U(1)\) gauge field \(\tilde{a}\) with a coupling of \(g_a = g/\sqrt{2NM}\). The gauge coupling being \(g_a \propto g_h/\sqrt{M}\) is what leads to the fact that leading order corrections to the spin-susceptibility come from RPA diagrams of the boson \(\tilde{h}\), instead of gauge corrections to the fermion propagator and the vertex. Note that because of spin-sums, RPA gauge corrections to the spin susceptibility vanish to all orders.\par
Now, in the symmetry broken phase, the Weiss fields have a non-zero expectation value around which we can consider excitations \(\tilde{h}^a(k) = \tilde{h}_0^a \d(k)+\d \tilde{h}^a(k)\). The uniform condensate part will give the fermions a matrix mass as \((1/\sqrt{N})\bar{\psi}_{\a, \ell}(k)\tilde{h}_0^a T^a_{\a\b}\psi_{\b, \ell}(k)\). The full low-energy action in the ordered phase hence looks like (upto a total constant corresponding to the energy of the condensate)
\begin{equation}
    \label{action large NM ordered}
    \begin{split}
        \ms = &\int d^3k\:\left[ \bar{\psi}_{\a, \ell}(k)\left(\slashed{k}\d_{\a\b}-\frac{1}{\sqrt{N}}\tilde{h}_0^aT^a_{\a\b}\right)\psi_{\b, \ell}(k)+\frac{1}{4}\tilde{F}^2+\d\tilde{h}^a(k)\mj^{-1}(k)\d\tilde{h}^a(-k)\right]\\
        +&\int d^3k\:d^3q\:\left[\frac{1}{\sqrt{2N}}\d\tilde{h}^a(q)\bar{\psi}_{\a, \ell}(k+q)T^a_{\a\b}\psi_{\b, \ell}(k) + \frac{g}{\sqrt{2NM}}\bar{\psi}_{\a, \ell}(k+q)\slashed{\tilde{a}}(q)\psi_{\a, \ell}(k)\right]\,.
    \end{split}
\end{equation}
One can therefore integrate out the fermions altogether from this theory and get an effective action in terms of the Weiss field excitations \(\d\tilde{h}^a\) and the gauge boson \(a_{\m}\). Note that since \(\tilde{h}^a\) are not charged under \(a_{\m}\) (they are gauge invariant fields) so are \(\d\tilde{h}^a\). Hence, integrating out the fermions \emph{cannot} couple \(\d\tilde{h}^a\) with \(a_{\m}\) directly, it can only couple \(\d\tilde{h}^a\) to gauge invariant operators like \(F_{\m\n}\) (\textit{e.g.} \(\partial_{\m}\d\tilde{h}^a \partial_{\n}\d\tilde{h}^a F^{\m\n}\)). Therefore the effective action at leading order (quadratic) would entirely consist of decoupled sectors \(\ms_{eff}[\d\tilde{h}]+\ms_{eff}[a_{\m}]\). Now we know that once the charged fermions are gapped in a compact \(U(1)\) gauge theory in \(2+1\) dimensions, monopoles are allowed to proliferate \cite{polyakov}. Monopoles in our formulation correspond to \(2\pi\) local flux insertions of \(a_{\m}\) in Eq.\eqref{action large NM}. Subsequently, the photons are gapped and there are no residual gapless gauge excitations. The effective low energy action is simply that of the Weiss fields themselves \(\ms_{eff}[\d\tilde{h}]\), just like a conventional magnet. To leading order, it will schematically be
\begin{equation}
    \label{action large NM ordered}
    \begin{split}
        \ms_{eff}[\d\tilde{h}] = &\int d^3k\:\left[\d\tilde{h}^a(k)\left(\d^{ab}\mj^{-1}(k)-\frac{M}{N}\c_0^{ab}(k)\right)\d\tilde{h}^b(-k)\right]+\mo(\d\tilde{h}^4)\,,
    \end{split}
\end{equation}
where \(\c_0^{ab}(k)\) is the mean-field spin susceptibility coming from the fermion bubble diagram per flavor \(\ell\) in \(SU(M)\).

\section{VMC} \label{appendix:VMC}

In our VMC implementation, the auxillary parton Hamiltonian we have considered looks like
\begin{equation}
    \label{vmc aux ham}
    \begin{split}
        \mh^{f} = -\sum_{\expval{i, j}_1}t_{ij}\left(f^{\dagger}_{i, \a}f_{j, \a}+\mathrm{h.c.}\right)-\frac{1}{2}\sum_{i}h_i^a f^{\dagger}_{i, \a}\s^a_{\a\b}f_{i, \b}\,,
    \end{split}
\end{equation}
where \(|t_{ij}|=1\) can be held fixed. The site-dependent Weiss fields \(h_i^a\) depend on the specific ordering pattern being considered. In this paper we worked with a \(L=4\times4\times6\) triangular lattice with an enlarged unit cell of \(6\) sites to accommodate (a) \(\bQ=\bK\) ordering like the \(120^{\circ}\) coplanar, or the supersolid ordering, (b) \(\bQ=\bM\) ordering like the stripe order, and (c) a \(\pi\)-flux state. The co-planar and collinear orders can be captured by a single Weiss field strength \(h\) whereas the supersolid order requires three independent Weiss fields to keep track of. \par

\begin{figure}[!ht]
 \centering
    \includegraphics[width=0.5\textwidth]{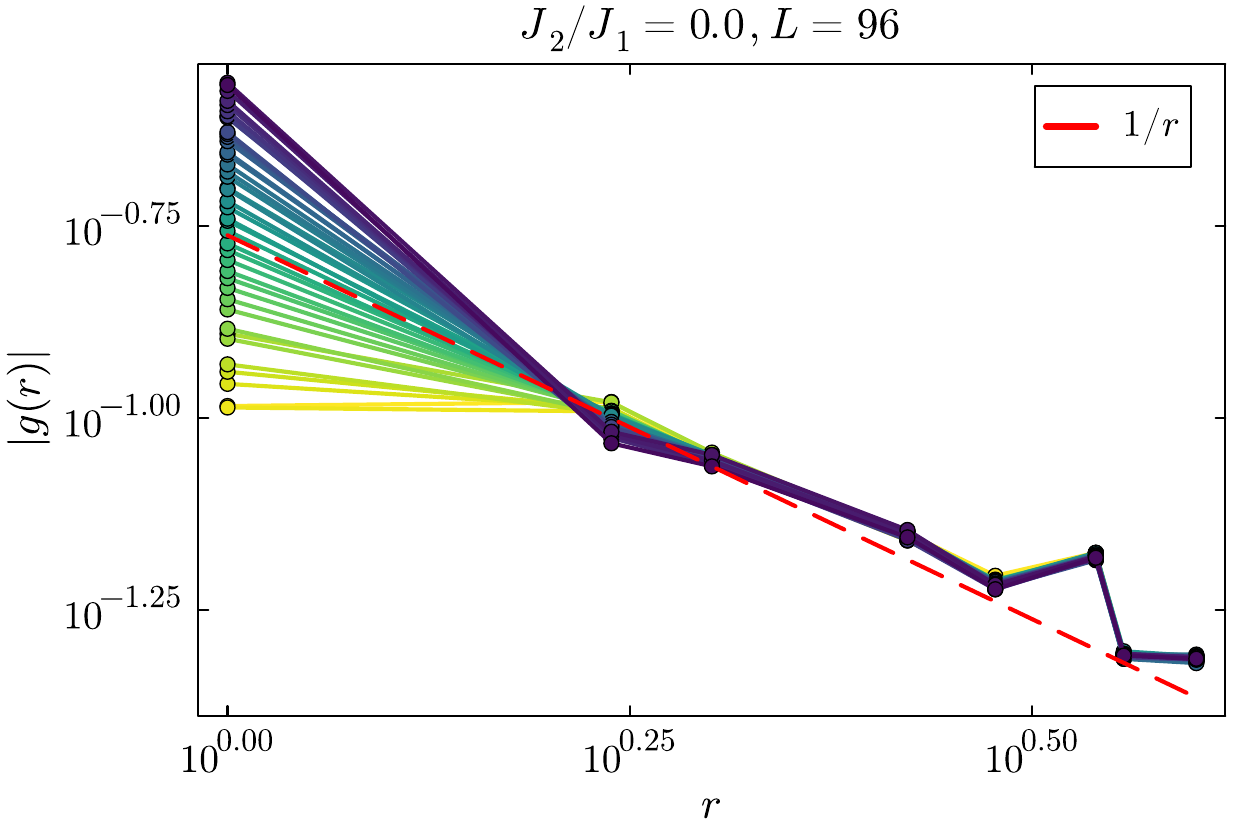}
 \caption{ The Jastrow factors \(g(r)\) in the optimized solutions of the supersolid phase at \(J_2/J_1=0\) which follow a \(\sim 1/r\) power law. The different color correspond to different strengths of the anisotropy.}
\label{fig:VMC jastrows}
\end{figure}

The mean-field ground state \(\ket{\psi_{\mathrm{MFT}}}\) can then be found by diagonalizing this quadratic Hamiltonian and filling states upto half-filling. This mean-field state variational state is then Gutzwiller projected to get the full many-body spin wavefunction \(\ket{\psi} = \ket{\psi(\{t_3, h_i, g_{ij}\})} = \exp\left(-\sum_{i, j} g_{ij} \hat{S}_i^z \hat{S}_j^z\right)\hat{P}\ket{\psi_{\mathrm{MFT}}(t_3, h_i)}\), where \(g_{ij}\) are long-range (upto \(10^{th}\) nearest-neighbor) Jastrow factors to further enhance spin-correlations. We then measure the expectation value of the full spin Hamiltonian in Eq.\eqref{easy plane Hamiltonian full} as \(E(\{h_i, g_{ij}\}) = \expval{\psi | \mh^{\ms} |\psi}/\expval{\psi|\psi}\) by sampling the many-body wavefunction through 10,000 (thermalization) + 10,000 (measurement) Monte Carlo sweeps. This energy is then optimized \textit{w.r.t} the variational parameters of the wavefunction using stochastic reconfiguration gradient descent \cite{becca_sorella_2017} for 250 steps. The optimized Jastrow factors in the supersolid ordered phase are plotted in Fig.\ref{fig:VMC jastrows} 

\section{Mean-field DSL properties}\label{appendix:MFT}
\begin{figure}[!ht]
 \centering
    \includegraphics[width=0.25\textwidth]{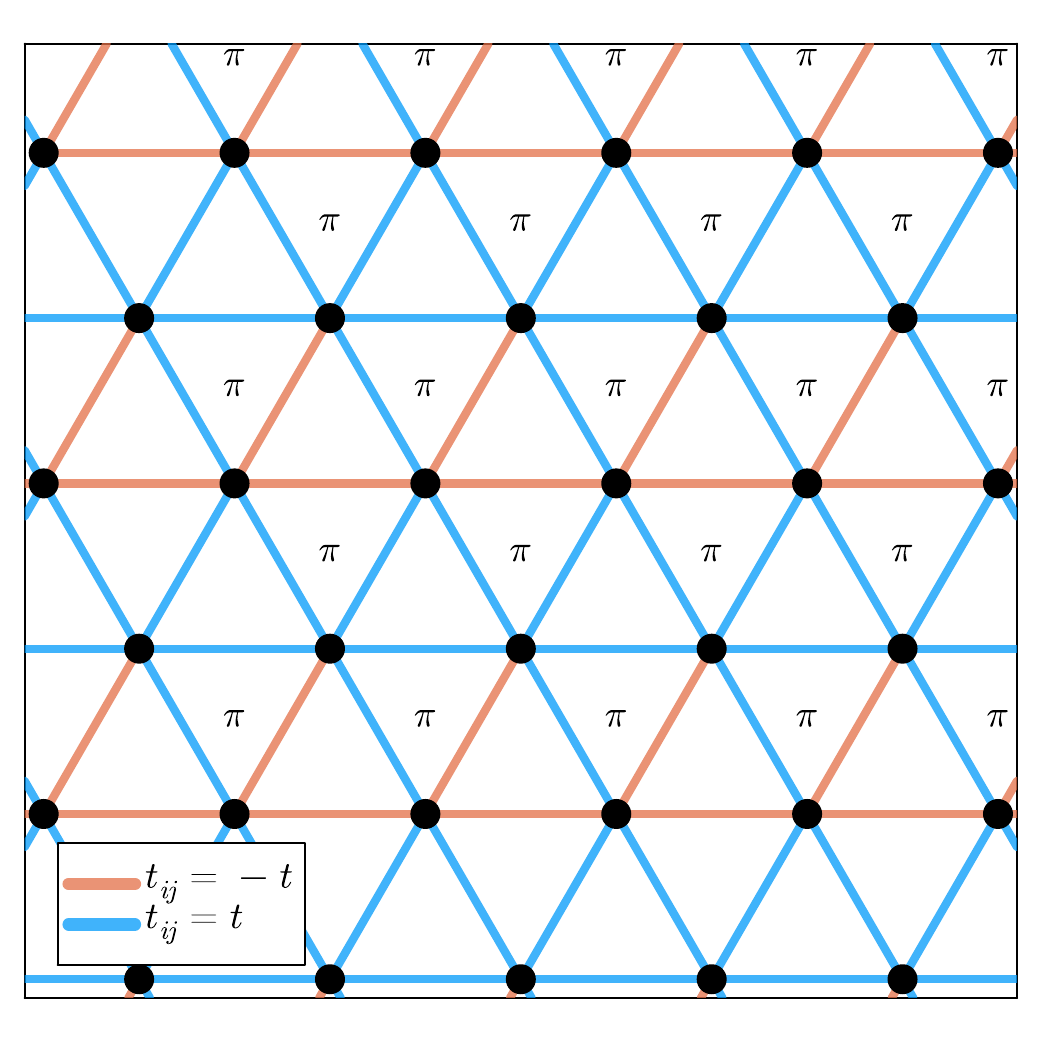}
 \caption{ The triangular lattice \(\pi\)-flux state with alternating fluxes of \(0\) and \(\pi\) in neighboring triangular plauquettes. We have also shown one possible gauge choice for the hoppings consistent with this flux state.}
\label{fig:pi flux hoppings}
\end{figure}
\begin{figure}[!ht]
 \centering
    \includegraphics[width=0.29\textwidth]{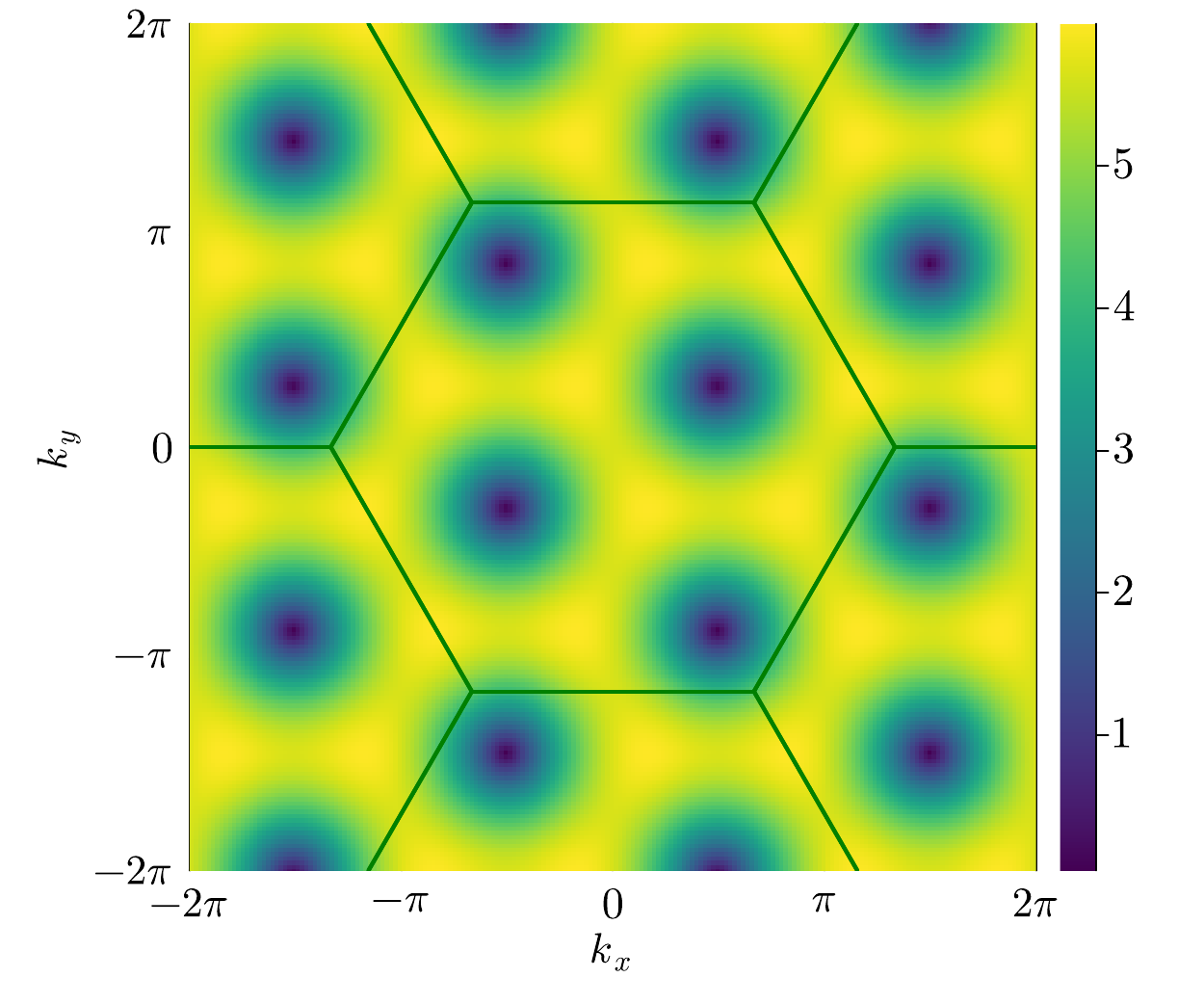}
    \includegraphics[width=0.5\textwidth]{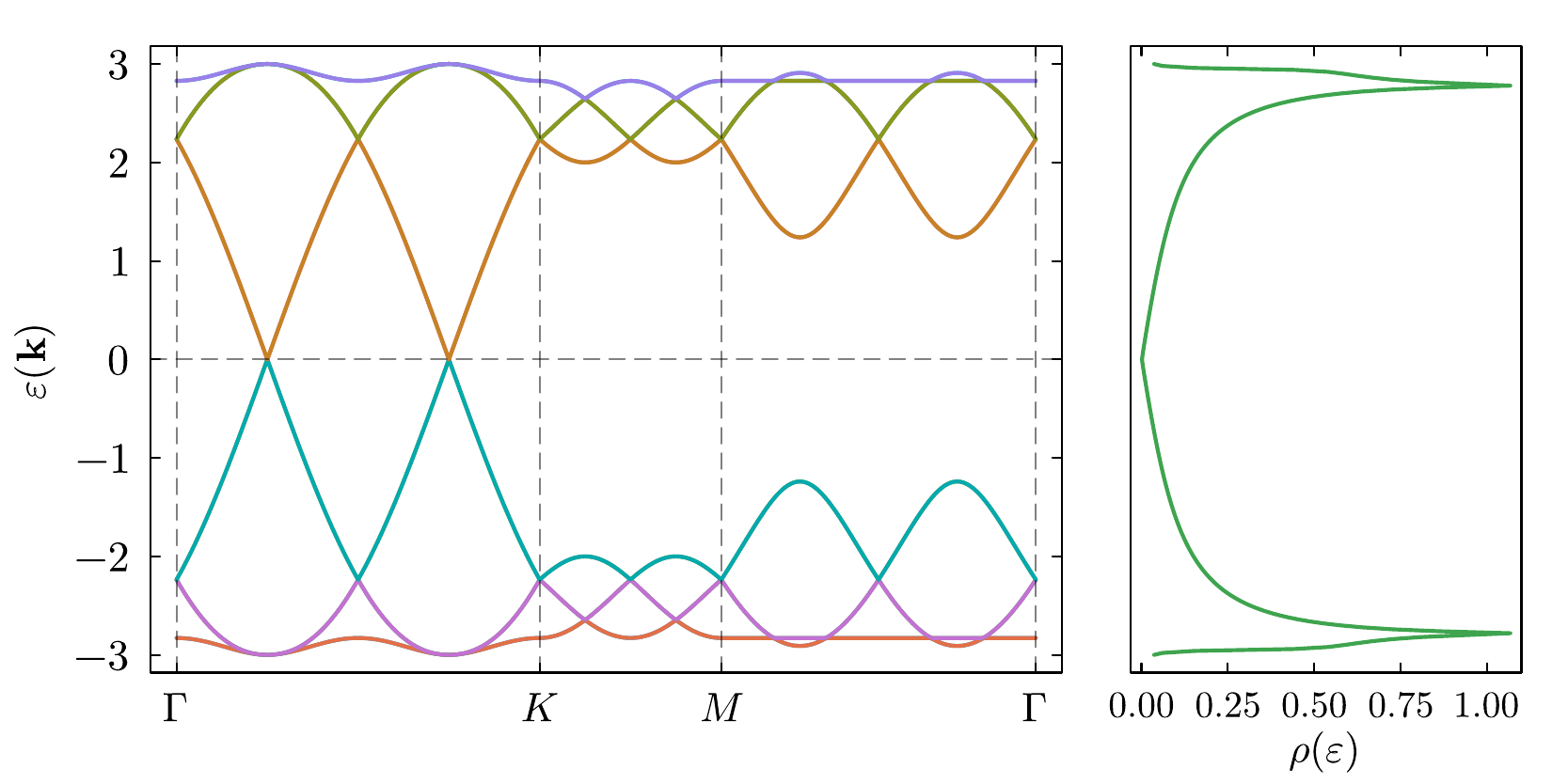}
 \caption{ (Left) The triangular lattice \(\pi\)-flux shows two doubly-degenerate Dirac nodes. The two nodes are related by inversion at \(\pm M/2\) in the reduced Brillouin zone, and the double degeneracy comes from spin. The resulting nesting vectors for the nodes are all the \(M\) points of the original triangular Brillouin zone. (Right) The band structure at the mean field level for a \(\pi\)-flux state with Dirac nodes. We also plot the corresponding density of state \(\r(\w)\) (in green). Note that the unit cell has been enlarged to 6 sites (to accommodate ordering Weiss fields) and hence the location of the nodes move when compared to Fig.\ref{fig:pi flux nodes}.}
\label{fig:pi flux nodes}
\end{figure}

Here we mention a few details about the mean-field state for the DSL and supersolid phase. The DSL is realized through the so called \(\pi\)-flux state where each spin of the parton sees a gauge flux of \(0\) or \(\pi\) in the up and down triangles of the lattice. This state as shown in Fig.\ref{fig:pi flux hoppings} does not break any physical symmetries (the \(\pi\) and \(0\) flux can be interchanged through a particle-hole transformation on the partons) like translation or rotation. However, the parton hoppings naively seem to break some symmetries in a \emph{fixed} gauge choice and requires doubling the unit cell to two sublattices. The resulting Dirac nodes shown in Fig.\ref{fig:pi flux nodes} in the band structure are related by inversion, and doubly degenerate due to spin-\(\ua\) and spin-\(\da\).

\begin{figure}[!ht]
 \centering
    \includegraphics[width=0.48\textwidth]{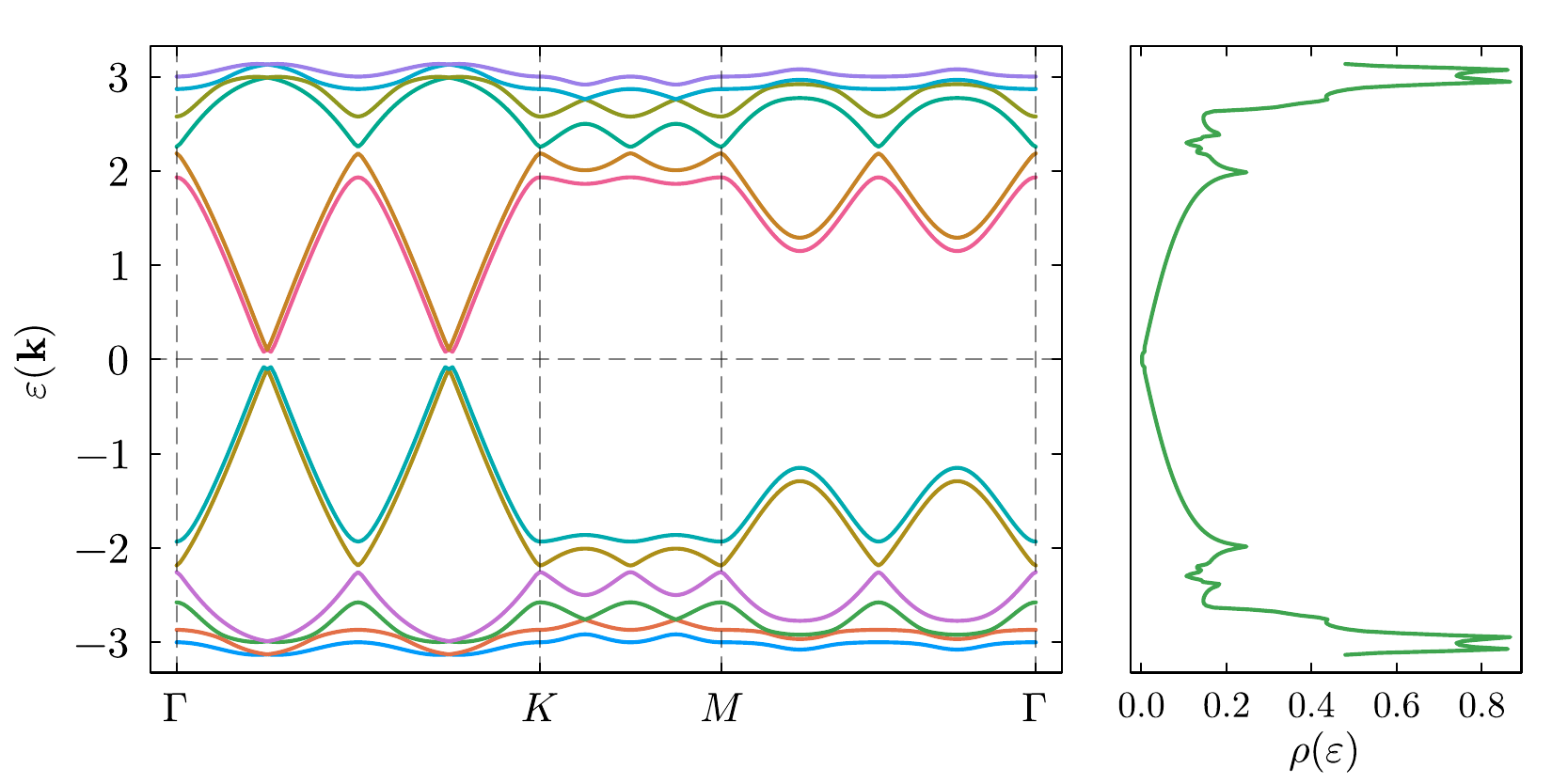}
    \includegraphics[width=0.48\textwidth]{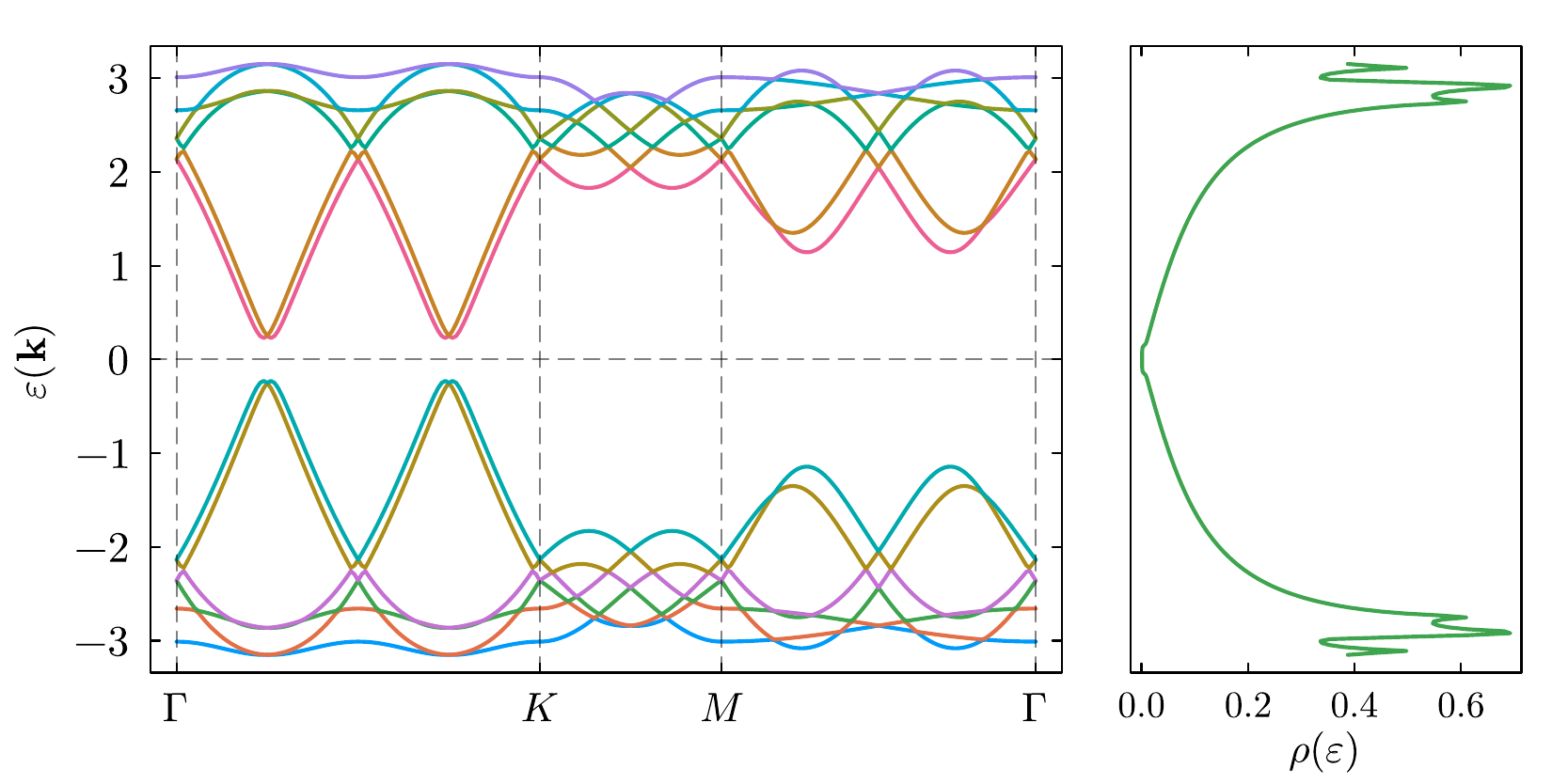}
 \caption{ The band structure at the mean field level for (a) a \(\pi\)-flux state along with supersolid Weiss fields which gap out the nodes and split the spin-degeneracy, and (b) a \(\pi\)-flux state along with stripe Weiss fields which also gap out the nodes. For each case we have also shown the corresponding density of state \(\r(\w)\) (in green).}
\label{fig:ordering DOS}
\end{figure}

To accommodate the supersolid order and stripe order on top of the DSL, one needs to enlarge the unit cell further to hold 6 sublattices. The supersolid order gaps out the Dirac nodes through its in-plane component, while also splits spin-degeneracy because of its out-of-plane component. The stripe order also gaps out the Dirac nodes. The band structures and corresponding density of states are shown in Fig.\ref{fig:pi flux nodes} and Fig.\ref{fig:ordering DOS}.

\end{document}